\documentclass[letterpaper,twocolumn,10pt]{article}
\PassOptionsToPackage{hyphens}{url}\usepackage{hyperref}
\usepackage{usenix2026}
\usepackage{graphicx}
\usepackage{tikz}
\usepackage{amsmath}
\usepackage{amssymb}
\usepackage{booktabs}
\usepackage{cleveref}
\usepackage{paralist}
\usepackage{filecontents}
\newcommand{\para}[1]{\smallskip\noindent\textbf{{#1.}}}
\usepackage{hyperref}
\usepackage{minted}
\usepackage[utf8]{inputenc}    
\usepackage{enumitem}    
\usepackage{comment}
\usepackage{tabularx}
\usepackage{subcaption}
\usepackage{hyperref} 
\usepackage{makecell}
\usepackage{threeparttable}
\usepackage[table]{xcolor}
\crefname{algocf}{algorithm}{algorithms}
\Crefname{algocf}{Algorithm}{Algorithms}

\usepackage[ruled,vlined,linesnumbered]{algorithm2e}    
\usepackage{algpseudocode}
\usepackage{xurl} 
\usepackage{etoolbox}
\newbool{extendedversion}
\setbool{extendedversion}{true} 
\newcommand{\ifextended}[1]{\ifbool{extendedversion}{#1}{}}
\newcommand{\ifmain}[1]{\ifbool{extendedversion}{}{#1}}

\newif\ifanonymous

\anonymousfalse

\newcommand{\anon}[2]{%
  \ifanonymous%
    #1%
  \else%
    #2%
  \fi%
}

\newcommand{\formatdoi}[1]{\href{https://doi.org/\detokenize{#1}}{#1}}

\newcommand{\ignore}[1]{}
\newcommand{\asm}[1]{\texttt{#1}}
\newcommand{\sys}[1]{\texttt{#1}}
\newcommand{\kw}[1]{\textit{#1}}
\newcommand{\kwb}[1]{\textbf{#1}}

\newenvironment{CompactItemize}%
  {\begin{list}{$\blacktriangleright$}%
    {\leftmargin=\parindent \itemsep=2pt \topsep=2pt
     \parsep=0pt \partopsep=0pt}}%
  {\end{list}}

\newcommand{\malloc}{{\texttt{Malloc}}}
\newcommand{\linklist}{{\textsf{Linked-List}}}

\newcommand{\traceoverall}{$2-3\times$\xspace}

\newcommand{\traceusersignpixel}{$1.7\times$\xspace}
\newcommand{\traceuserverifypixel}{$11\times$\xspace}
\newcommand{\tracekernelsignpixel}{$1.7\times$\xspace}
\newcommand{\tracekernelverifypixel}{$8\times$\xspace}

\newcommand{\tracemprotsignpixel}{$2.9\times$\xspace}

\newcommand{\tracemprotverifypixel}{$22.5\times$\xspace}

\newcommand{\specworstpixeleight}{\glibcsynchmmerx}
\newcommand{\glibcsynchmmerx}{$6.64\times$\xspace}
\newcommand{\glibcsynchrefx}{$2.37\times$\xspace}
\newcommand{\glibcasyncgccbig}{$1.82\times$\xspace}
\newcommand{\glibcasyncomnetbig}{$1.54\times$\xspace}
\newcommand{\glibcasyncxalancbig}{$1.54\times$\xspace}

\newcommand{\glibcasyncbigmax}{$1.82\times$\xspace}

\newcommand{\bufcopyPl}{$1.079\times$\xspace} 
\newcommand{\buflockPl}{$1.01\times$\xspace} 
\newcommand{\bufcopyPb}{$1.118\times$\xspace} 
\newcommand{\buflockPb}{$1.031\times$\xspace} 
\newcommand{\bufcopyPx}{$1.132\times$\xspace} 
\newcommand{\buflockPx}{$1.415\times$\xspace} 
\newcommand{\bufcopyPIXl}{$1.071\times$\xspace} 
\newcommand{\buflockPIXl}{$1.037\times$\xspace} 
\newcommand{\bufcopyPIXb}{$1.104\times$\xspace} 
\newcommand{\buflockPIXb}{$1.085\times$\xspace} 
\newcommand{\bufcopyPIXx}{$1.103\times$\xspace} 
\newcommand{\buflockPIXx}{$1.381\times$\xspace} 
\newcommand{\bufcopyampere}{$1.097\times$\xspace} 
\newcommand{\buflockampere}{$1.051\times$\xspace} 

\newcommand{\buflockovercopymax}{$1.087\times$\xspace} 

\newcommand{\polybenchlittlemax}{$1.31\times$\xspace}
\newcommand{\polybenchlittlegeomean}{$1.04\times$\xspace}

\newcommand{\polybenchbigmax}{$1.26\times$\xspace}
\newcommand{\polybenchbiggeomean}{$1.03\times$\xspace}

\newcommand{\polybenchxmax}{$3.65\times$\xspace}
\newcommand{\polybenchxgeomean}{$1.95\times$\xspace}
\newcommand{\polybenchamperemax}{$5.6\times$\xspace}

\newcommand{\tagsslittlegeomean}{$1.023\times$\xspace}
\newcommand{\tagssbiggeomean}{$1.028\times$\xspace}
\newcommand{\tagssxgeomean}{$1.026\times$\xspace}
\newcommand{\tagssallcoresgeomeanmin}{$1.023\times$\xspace}
\newcommand{\tagssallcoresgeomeanmax}{$1.028\times$\xspace}
\newcommand{\tagsslittlemax}{$1.13\times$\xspace}

\newcommand{\rallittlegeomean}{$1.034\times$\xspace}
\newcommand{\ralbiggeomean}{$1.164\times$\xspace}
\newcommand{\ralxgeomean}{$1.335\times$\xspace}

\newcommand{\vttlittlegeomean}{$1.013\times$\xspace}

\newcommand{\vttxgeomean}{$1.006\times$\xspace}
\newcommand{\vttlittlemax}{$1.05\times$\xspace}

\newcommand{\ifctpovraybig}{$1.12\times$\xspace}
\newcommand{\ifctpovrayx}{$1.08\times$\xspace}

\newcommand{\ifctbiggeomean}{$1.04\times$\xspace}

\newcommand{\amperehmmer}{$1.43\times$\xspace}

\newcommand{\ampereastar}{$1.10\times$\xspace}
\newcommand{\ampereastarstldisabled}{$1.20\times$\xspace}
\newcommand{\amperespecgeomean}{$1.10\times$\xspace}
\newcommand{\amperespecgeomeanstldisabled}{$1.05\times$\xspace}
\newcommand{\amperespecseventeengeomean}{$1.03\times$\xspace}
\newcommand{\amperespecnostlgeomean}{$1.05\times$\xspace}
\newcommand{\amperejacobinative}{$8.8\times$\xspace} 

\newcommand{\ampererocksdbmaxdrop}{$1.18\times$\xspace} 

\newcommand{\amperememcachelargedropmax}{$1.40\times$\xspace}

\newcommand{\amperefixmax}{$1.13\times$\xspace}

\newcommand{\amperefixmemcached}{$1.05\times$\xspace}

\newcommand{\amperegccselectivebefore}{$1.23\times$\xspace}
\newcommand{\amperegccselectiveafter}{$1.15\times$\xspace}
\newcommand{\peightlittlegccselectivebefore}{$1.26\times$\xspace}
\newcommand{\peightlittlegccselectiveafter}{$1.15\times$\xspace}

\newcommand{\applespecseventeengeomean}{$1.02\times$\xspace}

\usepackage[available]{usenixbadges}
\begin{document}

\def\ShortName{\textsc{Tag}}

\title{ARM MTE Performance in Practice\ifextended{ (Extended Version)}}

\anon{
\author{Anonymized for submission}
}{
\author{
{\rm Taehyun Noh}\\ UT Austin \and
{\rm Yingchen Wang}\\ UC Berkeley \and
{\rm Tal Garfinkel}\\ Google \and
{\rm Mahesh Madhav}\\ Ampere Computing \and
{\rm Daniel Moghimi}\\ Google \and
{\rm Mattan Erez}\\ UT Austin \and
{\rm Shravan Narayan}\\ UT Austin}
}

\newcommand\yw[1]{\textcolor{blue}{yingchen: #1}}
\newcommand\daniel[1]{\textcolor{purple}{daniel: #1}}
\newcommand\taehyun[1]{\textcolor{teal}{taehyun: #1}}
\newcommand\talg[1]{\textcolor{red}{talg: #1}}

\maketitle
\begin{abstract}

We present the first comprehensive analysis of ARM MTE hardware performance on four different microarchitectures: ARM Big (A7x), Little (A5x), and Performance (Cortex-X) cores on the Google Pixel 8 and Pixel 9, and on Ampere Computing's AmpereOne CPU core. We also include preliminary analysis of MTE on Apple's M5 chip. We investigate performance in MTE's primary application---probabilistic memory safety---on both SPEC CPU benchmarks and in server workloads such as RocksDB, Nginx, PostgreSQL, and Memcached. While MTE often exhibits modest overheads, we also see performance slowdowns up to \specworstpixeleight on certain benchmarks. We identify the microarchitectural cause of these overheads and where they can be addressed in future processors. We then analyze MTE's performance for more specialized security applications such as memory tracing, time-of-check time-of-use prevention, sandboxing, and CFI. In some of these cases, MTE offers significant advantages today, while the benefits for other cases are negligible or will depend on future hardware. Finally, we explore where prior work characterizing MTE performance has either been incomplete or incorrect due to methodological or experimental errors.
\end{abstract}

\section{Introduction}

Memory safety vulnerabilities remain one of the biggest challenges in systems security. Consequently, ARM's memory tagging extension (MTE)~\cite{armmte}---that detects and probabilistically mitigates these bugs---has been highly anticipated~\cite{pzeromte,cisa-case-for-safety}. MTE promises several important properties including: \kw{generality}---detecting both temporal and spatial safety bugs; \kw{compatibility}---the ability to work with unmodified applications and only minimal changes to existing allocators and operating systems; and \kw{low overhead}---allowing it to be on by default on production settings.
The reality of this last promise, the actual cost of MTE in it different modes (SYNC, ASYNC, and ASYMM), can only be found in silicon. Understanding that reality is our goal in this paper.

Security features live and die by their performance. The choice to use any feature is a cost-benefit analysis; the higher the cost, the less likely a feature is to see production use. At present, there is a significant gap in our knowledge of MTE's performance costs in real hardware; what those costs are for different use cases; where those costs come from at the microarchitectural
level; how those costs vary across different microarchitectures; and which costs are fundamental vs.\ fixable in the next generation of hardware.

To bridge this gap, we analyze the performance of MTE in five different microarchitectures, ranging from mobile (Pixel 8 and 9 ``Performance'' (Cortex-X), ``Big'' (A7x), and in-order ``Little'' (A5x)) cores
\footnote{We technically measured 6 different Pixel cores---3 on the Pixel 8, and 3 on the Pixel 9. But both SoCs exhibit similar performance patterns.}
to laptop (Apple M5~\cite{apple-mie}) and high-end servers (AmpereOne~\cite{ampereone}).
We evaluate these architectures using traditional and server benchmarks across a variety of scenarios. We analyze performance for both the primary use case of probabilistic memory safety and several advanced applications of MTE inspired by prior work~\cite{greathouse2012case, shastri2024hmtrace, chen2024limitations, segue-cg, clang-cfi, cpi}, including time-of-check time-of-use (TOCTOU) attack mitigations, memory tracing, sandboxing, and control-flow integrity.
Finally, we carefully investigate performance regressions with different MTE modes and use targeted microbenchmarks and performance analysis tools to attribute those to microarchitecture decisions in different core types or software issues.

\para{Probabilistic memory safety}
We measure MTE's overhead in its primary use case---probabilistic memory safety---using SPEC CPU 2006 INT \cite{spec2006} (\S\ref{sec:perf}) and a collection of popular server workloads including RocksDB, Nginx, PostgreSQL, and Memcached~\cite{rocksdb, nginx, postgresql, memcached}.

Our main takeaway is that while many benchmarks run with low performance overheads, each microarchitecture exhibits outlier performance regressions. On SPEC INT 2006, these can reach up to \specworstpixeleight (Pixel Performance core with MTE SYNC) and be above \glibcasyncbigmax even in the more relaxed security guarantees of MTE ASYNC (Pixel Big core). The AmpereOne core also exhibits a wide range of performance overhead that is mostly in the single digits, but can reach \amperehmmer. A similar range of overheads is observed on the server benchmarks with AmpereOne.

\para{Microarchitectural analysis}
We investigate the cause of these performance cliffs in MTE's SYNC and ASYNC modes and found that they stem from some limitations in current microarchitectures:

\kw{Store serialization:} The Pixel 8 and 9 Performance cores exhibit significant performance cliffs of up to \specworstpixeleight on specific SPEC CPU benchmarks. We use targeted microbenchmark experiments to attribute these overheads to a design decision that serializes stores in MTE SYNC. This choice is limited to the Performance core.

\kw{Structural hazard on tag-checking of loads:} The Pixel 8 and 9 Big core exhibits slowdowns of up to \glibcasyncbigmax in both SYNC and, more importantly, ASYNC on certain SPEC CPU benchmarks. We used microbenchmarks to narrow down the cause of this overhead to bottlenecks in the structures that support in-flight tag-checks, which artificially restrict the number of allowed
in-flight memory operations.

\kw{Store-to-load forwarding:} AmpereOne had overheads on specific benchmarks up to \amperehmmer. We identified that this was because its store-to-load forwarding---an important optimization in out-of-order cores that forwards the result of older stores to younger loads without passing through the cache---appears to have inconsistent behavior.

On a positive note, when we shared our findings with Ampere, they
confirmed the presence of a performance issue in the store-to-load forwarding mechanism when interacting with MTE tag checks. They also mentioned that this has been fixed in the next-generation silicon supporting MTE. This provides strong support for the idea that MTE's
current limitations are surmountable and that an MTE implementation with consistently low overheads is possible and on the horizon. 

\para{Other sources of overheads} While overheads on server benchmarks were usually under \ampererocksdbmaxdrop, some memcached workloads slowed down by up to \amperememcachelargedropmax (\S\ref{subsec:server}).  Due to the Linux kernel's reliance on an ambiguous portion of ARM's specification, tagging overheads were imposed on all kernel memory operations---rather than just user space pages with tagging enabled---on Ampere chips. We shared our findings with Ampere and together we developed a fix, following suggestions from Linux kernel developers~\cite{lkml-mte-patch}.


\para{Beyond memory safety}
Next, we turn our attention to MTE's performance for other use cases. We developed MTE-based tools for other important uses of tagged architectures including: time-of-check time-of-use prevention based on buffer revocation (\S\ref{sec:toctou}), memory tracing (\S\ref{sec:trace}), optimizing sandboxing (\S\ref{sec:colorguard}), optimizing/enforcing CFI (\S\ref{sec:cfi}).

For memory tracing, we found that MTE enables simple implementation and low overhead, with performance roughly \traceoverall faster than other approaches like using page permissions.
For TOCTOU prevention, MTE offers performance boosts of up to \buflockovercopymax over alternate approaches like copying on most cores (although Pixel's Performance core alone shows a pessimization when using this technique).
In contrast, we find using MTE to enforce sandboxing or CFI offers no clear
performance benefits over existing software approaches~\cite{lfi,clang-cfi}; in
particular for sandboxing, where its performance cliffs could induce high
and hard to predict overheads.

\para{Other contributions}
We also explore how the performance overheads we observed are not captured by prior estimates of MTE's performance using analogs (\S\ref{subsec:analogs}), and sometimes due to experimental error (\S\ref{subsec:cage}), highlighting the importance of accurate evaluations on real hardware to provide a solid foundation for other work to build on.

We offer a few observations (\S\ref{sec:wishlist}) on how ARM could improve future standards to enhance MTE's functionality, e.g., enhancing MTE's value as a tracing mechanism, and improving ASYNC mode to provide sufficient information to diagnose MTE traps.

Finally, we note that while the microarchitecture implementations of MTE differ in design and performance characteristics, all seem to have encountered similar challenges, evidenced by performance degradation and cliffs on overlapping benchmark sets. This is true even for the original SPARC ADI implementation which struggles on the same benchmarks~\cite{kostya-tagging}. Thus, future implementations---e.g., ARM based CPU's from cloud providers~\cite{aws_graviton,google_axion} and
RISC-V implementations~\cite{riscvmte}---should be aware of these issues.

\section{Why MTE?}
\label{sec:bg}


ARM's Memory Tagging Extension (MTE)~\cite{armmte} is unique among current memory safety mitigations. It promises to detect and mitigate (with the caveats below) a wide range of temporal and spatial memory-safety bugs with no application changes\footnote{MTE only requires small changes to the OS and user space allocator.}, low enough overhead to be deployed in production, and a modest enough cost in silicon to make adoption feasible.

The key challenge of enforcing memory safety is tracking metadata, e.g., array bounds for spatial safety. Precisely tracking this state in hardware can be very challenging, and existing solutions often compromise performance, complexity, or compatibility~\cite{watson2015cheri,intel-mpx}. MTE avoids many of these compromises using a simpler approach to tracking metadata, which sacrifices precision but retains these other useful properties.

MTE relies on 4-bit tags, which map all allocations into 16-congruence classes. Upon allocation, the allocator assigns a random tag to the address range of each allocation and returns a pointer with the top bits containing a matching tag. Thus, if that pointer is used to access an object with a different tag (e.g., an out-of-bounds memory access, or a use of a dangling pointer), there is a 15/16 chance it will trap.

Allocators also employ techniques to improve MTE's detection of memory safety errors by reserving a unique tag for allocator data-structures or employing guard regions after large allocations~\cite{partap2022memory}, however, these techniques do not change the fact that MTE offers only probabilistic defenses against memory safety bugs.

MTE's probabilistic nature and other limitations, such as its reliance on tag secrecy to prevent pointer foraging, its vulnerability to tag leaks via side channels, and its lack of intra-object safety, have led to the perception that MTE is a debugging rather than a security feature.

However, this is misleading. Although MTE does not provide sound memory safety, it is very likely to render a great majority of memory safety-based exploits unreliable and detectable. Both of these are powerful properties.

Unreliable exploits are far less valuable to attackers. Imagine if spyware that relies on zero-click exploits to deploy—such as NSO's Pegasus~\cite{citizenlab-nso-whatsapp}—instead of silently compromising an application, 
caused it to repeatedly crash. Not only would this alert victims, but the subsequent crash report would also signal the bug's presence to both the application developer and platform vendor, 
likely significantly shortening the life of the zero-day exploit used.

Moreover, high-end vulnerabilities of this type are already worth millions of dollars~\cite{opzero-prices}; businesses that sell offensive tools that rely on them are only viable because they can often amortize this cost across many uses. If most of the memory safety vulnerabilities on targeted devices were rendered unreliable and rapidly detected and remediated, it could significantly change the cost of carrying out such attacks~\cite{dowd-zeroday}.

As MTE can detect the immediate causes of intrusions, it could also potentially enable more reliable detection and faster remediation of server-side bugs. Thus, while MTE is not a substitute for sound mitigations for memory safety bugs, it could significantly shift the battlefield of the "eternal war in memory"~\cite{eternal-war} toward the defender's favor.

\subsection{How MTE Works} \label{sec:mte}

\begin{figure}[t]
\centering
\includegraphics[width=1\linewidth]{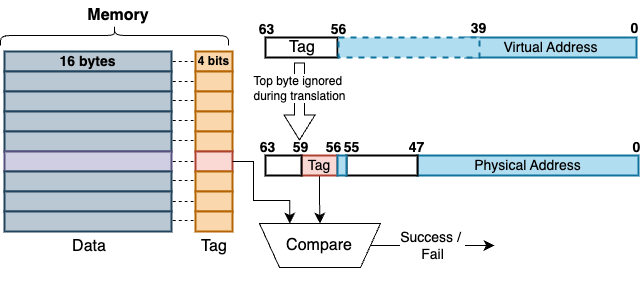}
\caption{MTE overview. MTE allows assigning of 4-bit tags to each 16-byte memory granule. Subsequent reads from (or writes to) tagged memory must have the correct tag in bits[56:59] of the virtual address. An incorrect tag results in a fault in MTE SYNC, or a deferred fault in MTE ASYNC.}
\label{fig:mte}
\end{figure}

MTE is a form of tagged memory~\cite{tag-advantages,TAG-guide,witchel2002mondrian,weiser2019timber,watson2015cheri,de2015micro}, i.e., it associates metadata (a tag) with each portion of the processor's address space. In MTE's case, this metadata is a 4-bit tag (color) for each 16-byte range (granule) of physical memory. The high-level operation of checking tags is shown in Figure~\ref{fig:mte}. This design is based on the Application Data Integrity (ADI)---the memory tagging feature in the SPARC M7~\cite{adi-m7} that used 64 byte granules. MTE's smaller granule size reduces the alignment constraint that ADI's larger granules impose on allocators, resulting in less wasted memory~\cite{kostya-tagging}.
%
MTE supports three different modes of operation that control when and how it
handles tag mismatches:

\para{SYNC mode} In SYNC mode, the processor takes a precise exception on a tag mismatch. Thus, it provides \kw{precise enforcement}---enabling access control; and \kw{precise reporting}---it  identifies the faulting instruction, making it ideal for diagnosing the cause of a fault. While SYNC mode is the most useful, it is also challenging to implement efficiently. This is because MTE's tag check adds an extra operation to each memory instruction; for example, a load becomes: load-tag, check-tag, load-value. In SYNC mode, the first two operations must complete before the final operation (the actual load) can commit. This additional work and the ordering constraint require careful modifications to several parts of the pipeline to support efficiently; as we see in \S\ref{subsec:mte-slowdowns}, implementations without careful optimization can add large overheads.

\para{ASYNC mode} In ASYNC mode, the dependency between the load-tag/check-tag and the memory operation (load) is relaxed. The only guarantee the processor provides is that a flag register will be set on a tag mismatch that the operating system can poll. By convention, the operating system checks this flag at the next system call and it immediately generates a fault if set. This convention implies that a mismatch will still be caught before there are externally visible side effects (modulo side channels and shared memory). Even so, ASYNC's lack of precision makes triaging crashes much harder---millions of crashes already go undiagnosed in production systems~\cite{retracer}—thus, it is less useful for bug finding or intrusion detection.

\para{ASYMM mode} ASYMM offers a trade-off between overhead and precision, using SYNC for loads and ASYNC for stores. However, for security, this is less than ideal, as memory corruption is generally caused by stores.

\para{Probabilistic Use} MTE was designed to probabilistically detect memory safety bugs. To enable this, the system allocator assigns a random color (1 of 16) to the memory of each allocation (e.g., a call to malloc()), and returns a pointer whose top byte has the same color as the allocation it refers to.

On every memory access (e.g., a load), the processor checks if the pointer's tag still matches the tag on the accessed memory. If the tag no longer matches, the processor takes an exception, either synchronously---to stop the access, or asynchronously---to detect it (\S\ref{sec:mte}). This exception implies that the pointer references the wrong object; either it is out of bounds (a spatial bug), or it is using an object that has been reallocated (a temporal bug). Thus, there is a 15/16 chance that a bug will be caught.

\para{Threat Model} The strength of this randomized tag approach depends on the attacker's model. For example, if an attacker can determine the tag of a particular region of memory, e.g., through a read gadget or a side-channel attack (e.g., Spectre)~\cite{tiktag, pzeromte}, and they have access to bugs that let them modify pointers, they can combine these elements to forge new pointers that can bypass MTE checks.

However, other applications of MTE (e.g., \S\ref{sec:flex}) do not necessarily depend on the secrecy of the tags. We consider analysis of side-channel attacks to be out-of-scope for this work.

\section{Contrasting MTE implementations}
\label{sec:mte-impls}
While the ARM ISA specifies MTE's interface, it leaves several details up to the implementer. We briefly describe two different MTE implementations we analyzed in this paper.

\subsection{Google Pixel MTE implementation}
\label{subsec:pixel-mte}

The Google Pixel 8 and 9 are based on the Tensor G3 and G4 processors, respectively. Each comes with three different types of cores: a low-power in-order "Little" core (A5x), a faster out-of-order "Big" core (A7x), and the fastest out-of-order "Performance" core (Cortex-X) \footnote{Specifically, \textbf{Pixel8}: A510, A715, and X3. \textbf{Pixel 9}: A520, A720, and X4.}. All cores support the different modes of MTE: SYNC, ASYNC, and ASYMM. To the best of our knowledge, these cores implement MTE using the reference implementation ARM provides to chip vendors.
Consequently, our results likely generalize to other devices that use ARM's reference cores. For example, the Samsung S23 relies on an almost identical configuration of Cortex-X, A7x and A5x cores as that of the Pixel 8. 

\para{Tag storage}
ARM ISA allows MTE implementations to flexibly choose how they store the tags associated with memory granules. In ARM's reference design, tag storage works by reserving a portion of the physical RAM exclusively for this purpose. Since MTE must store 4-bit tags for each 16 bytes of physical memory, this means that roughly 3\% of physical memory $(4/128)$ is reserved upfront for tag storage.
%

\para{Modified memory accesses and caches}
In ARM's implementation, fetching L1 and L2 data cache lines fetches the corresponding tags; i.e., each 64-byte cache line will also pull in 2 bytes of tags into the cache. If a miss occurs, the CPU performs two independent reads from memory to retrieve both the data and its associated tags. Data and tags are stored together in extended cache lines. The need to fetch additional data (i.e., tags) on a cache miss implies that ARM's design consumes more memory bandwidth when MTE is enabled.

\para{Impact of tag checks on the store pipeline}
Tag checks requires an important change in the store pipeline. Typically, stores are buffered in the store buffer (the CPU can asynchronously push these into main memory) and do not require cache lines to be retrieved during execution. However, with MTE SYNC enabled, stores must retrieve and check their associated tags before executing. Since tag fetches are coupled with data fetches, stores operate similarly to loads---fetching both data and tags into the cache during execution. This introduces a potential bottleneck that, if not handled with care, can significantly hinder application's performance (\S\ref{subsec:mte-slowdowns}).


\subsection{Ampere's MTE implementation}
\label{subsec:ampere-mte}
Ampere's AmpereOne CPUs are server-class ARM-compatible CPUs that offer up to 192 Ampere-1a cores~\cite{ampereone}. Ampere designs their own cores; thus, their performance profile is distinctly different from ARM's reference implementation. Ampere cores exclusively support SYNC mode.

\para{Tag storage}
Unlike ARM's reference design, Ampere has adopted co-located MTE tags with data across the storage hierarchy. Specifically, tags are stored in bits normally used for error-correction codes (ECC bits)~\cite{ampere-arxiv-v2} that are only accessible by the memory controller. This is particularly important in server environments where managing terabytes of physical memory is common---ARM's reference implementation, in contrast, would be harder to deploy in this environment, as reserving 3\% of physical memory results in a large reduction of usable memory.
Using ECC bits to store tags can potentially impose reductions on RAM reliability metrics. In its documentation\cite{ampere-arxiv-v2}, Ampere discusses an optimization they've used to minimize this reduction while supporting MTE.

\para{Modified memory accesses and caches}
Like the ARM reference implementation, Ampere CPUs also fetch data and tags together; these are stored in larger cache lines (extended from 64 bytes to 66 bytes), where the two extra bytes store tags. Unlike the ARM reference implementation, however, Ampere can retrieve data and associated tags with a single query to physical memory upon cache misses. Coherency also requires only a single transaction, compared to the reference implementation which requires two transactions. This means that cache misses do not impose any performance penalties or bandwidth reductions when MTE is enabled.

\para{Impact of tag checks on the store pipeline}
The impacts are similar to those in the ARM reference implementation.

\definecolor{low}{HTML}{306844} 
\definecolor{mid}{HTML}{FFFFFF}
\definecolor{high}{HTML}{c30101} 
\newcommand*{\opacity}{90}

\newcommand*{\minval}{0.0}
\newcommand*{\midval}{1.0}
\newcommand*{\maxval}{3}

\newcommand{\gradient}[1]{
    \ifdimcomp{#1pt}{>}{\maxval pt}{#1}{
        \ifdimcomp{#1pt}{<}{\minval pt}{#1}{
            \ifdimcomp{#1pt}{<}{\midval pt}{
                \pgfmathparse{int(round(100*(#1-\minval)/(\midval-\minval)))}
                \xdef\tempa{\pgfmathresult}
                \cellcolor{mid!\tempa!low!\opacity} #1
            }{
                \pgfmathparse{int(round(100*(#1-\midval)/(\maxval-\midval)))}
                \xdef\tempa{\pgfmathresult}
                \cellcolor{high!\tempa!mid!\opacity} #1
            }
            
    }}
}
\newcommand{\gradientbold}[1]{
    \ifdimcomp{#1pt}{>}{\maxval pt}{#1}{
        \ifdimcomp{#1pt}{<}{\minval pt}{#1}{
            \ifdimcomp{#1pt}{<}{\midval pt}{
                \pgfmathparse{int(round(100*(#1-\minval)/(\midval-\minval)))}
                \xdef\tempa{\pgfmathresult}
                \cellcolor{mid!\tempa!low!\opacity} \textbf{#1}
            }{
                \pgfmathparse{int(round(100*(#1-\midval)/(\maxval-\midval)))}
                \xdef\tempa{\pgfmathresult}
                \cellcolor{high!\tempa!mid!\opacity} \textbf{#1}
            }
            
    }}
}

\begin{figure*}[] 
\scriptsize
\centering

\setlength{\aboverulesep}{0pt} 
\setlength{\belowrulesep}{0pt}

\setlength{\tabcolsep}{3pt} 

\begin{minipage}[t]{0.72\textwidth}
\begin{tabularx}{0.975\textwidth}{l *{10}{c}}
\arrayrulecolor{gray}

& \multicolumn{3}{c}{\cellcolor[HTML]{f1a983}\textbf{Pixel 8 Perf Core}} 
& \multicolumn{3}{c}{\cellcolor[HTML]{f7c7ac}\textbf{Pixel 8 Big Core}} 
& \multicolumn{3}{c}{\cellcolor[HTML]{fbe2d5}\textbf{Pixel 8 Little Core}} 
& \cellcolor[HTML]{dbb3ff}\textbf{Ampere} \\

\textbf{SPEC CPU2006} 
& \textbf{ASYNC} & \textbf{SYNC} & \textbf{ASYMM} 
& \textbf{ASYNC} & \textbf{SYNC} & \textbf{ASYMM} 
& \textbf{ASYNC} & \textbf{SYNC} & \textbf{ASYMM} 
& \textbf{SYNC} \\

\midrule

              400.perlbench & \gradient{1.13} & \gradient{1.63} & \gradient{1.16} & \gradient{1.22} & \gradient{1.24} & \gradient{1.23} & \gradient{1.16} & \gradient{1.18} & \gradient{1.17} & \gradient{1.12} \\
              401.bzip2     & \gradient{1.08} & \gradient{1.84} & \gradient{1.16} & \gradient{1.00} & \gradient{1.40} & \gradient{1.00} & \gradient{1.01} & \gradient{1.10} & \gradient{1.01} & \gradient{1.14} \\
              403.gcc       & \gradient{1.21} & \gradient{1.57} & \gradient{1.29} & \gradient{1.82} & \gradient{1.83} & \gradient{1.81} & \gradient{1.19} & \gradient{1.26} & \gradient{1.20} & \gradient{1.23} \\
              429.mcf       & \gradient{1.05} & \gradient{1.19} & \gradient{1.05} & \gradient{1.12} & \gradient{1.17} & \gradient{1.12} & \gradient{1.07} & \gradient{1.07} & \gradient{1.07} & \gradient{1.01} \\
              445.gobmk     & \gradient{1.01} & \gradient{1.04} & \gradient{1.01} & \gradient{1.02} & \gradient{1.04} & \gradient{1.03} & \gradient{1.02} & \gradient{1.02} & \gradient{1.01} & \gradient{1.01} \\
              456.hmmer     & \gradient{1.00} & \cellcolor[HTML]{960000}\textcolor{white}{6.64}  & \gradient{1.12} & \gradient{1.00} & \gradient{1.01} & \gradient{1.00} & \gradient{1.02} & \gradient{1.05} & \gradient{1.02} & \gradient{1.43} \\
              458.sjeng     & \gradient{1.03} & \gradient{1.06} & \gradient{1.03} & \gradient{1.02} & \gradient{1.02} & \gradient{1.00} & \gradient{1.22} & \gradient{1.24} & \gradient{1.22} & \gradient{1.01} \\
              462.libquantum& \gradient{1.07} & \gradient{1.06} & \gradient{1.05} & \gradient{1.05} & \gradient{1.06} & \gradient{1.04} & \gradient{1.01} & \gradient{1.12} & \gradient{1.05} & \gradient{1.04} \\
              464.h264ref   & \gradient{1.00} & \gradient{2.37} & \gradient{1.01} & \gradient{1.00} & \gradient{1.08} & \gradient{1.00} & \gradient{1.03} & \gradient{1.14} & \gradient{1.03} & \gradient{1.07} \\
              471.omnetpp   & \gradient{1.16} & \gradient{1.38} & \gradient{1.18} & \gradient{1.54} & \gradient{1.60} & \gradient{1.51} & \gradient{1.10} & \gradient{1.20} & \gradient{1.11} & \gradient{1.13} \\
              473.astar     & \gradient{1.03} & \gradient{1.07} & \gradient{1.03} & \gradient{1.07} & \gradient{1.10} & \gradient{1.05} & \gradient{1.07} & \gradient{1.32} & \gradient{1.07} & \gradient{1.04} \\
              483.xalancbmk & \gradient{1.15} & \gradient{1.32} & \gradient{1.16} & \gradient{1.54} & \gradient{1.57} & \gradient{1.55} & \gradient{1.14} & \gradient{1.18} & \gradient{1.16} & \gradient{1.09} \\
              \midrule
              \textbf{Geomean} & \gradientbold{1.08} & \gradientbold{1.56} & \gradientbold{1.10} & \gradientbold{1.18} & \gradientbold{1.23} & \gradientbold{1.17} & \gradientbold{1.08} & \gradientbold{1.14} & \gradientbold{1.08} & \gradientbold{1.10} \\

\bottomrule
\end{tabularx}
\end{minipage}
\begin{minipage}[c]{0.247\textwidth}
\centering
\begin{tabularx}{\textwidth}{l *{2}{c}}
\arrayrulecolor{gray}

& \cellcolor[HTML]{dbb3ff}\textbf{Ampere}
& \cellcolor[HTML]{ffc5d3}\textbf{Apple M5} \\

\textbf{SPEC CPU2017} 
& \textbf{SYNC}
& \textbf{SYNC} \\

\midrule
    600.perlbench\_s &\gradient{1.02} &	\gradient{1.01} \\
    602.gcc\_s       & \gradient{1.03} &	\gradient{1.03} \\
    605.mcf\_s       & \gradient{1.08} &	\gradient{1.00} \\
    620.omnetpp\_s   & \gradient{1.03} &	\gradient{1.12} \\
    623.xalancbmk\_s & \gradient{1.14} &	\gradient{1.03} \\
    625.x264\_s      & \gradient{1.00} &	\gradient{1.02} \\
    631.deepsjeng\_s & \gradient{1.01} &	\gradient{1.02} \\
    641.leela\_s     & \gradient{1.01} &	\gradient{1.00} \\
    657.xz\_s        & \gradient{1.03} &	\gradient{1.03} \\
    \midrule
    \textbf{Geomean} & \gradientbold{1.03} &	\gradientbold{1.02} \\
    \midrule
    \\ 
    \midrule
    502.gcc\_r	     & \gradient{1.12}     &	\gradient{1.09} \\
    502.gcc\_r (input 5)	& \gradient{1.25}     &	\gradient{1.29} \\
\bottomrule
\end{tabularx}
\end{minipage}

\par
\caption{\textbf{Performance Cliffs Across MTE Impl. and Modes.}~\textit{MTE Overhead on SPEC CPU INT 2006 on Pixel 8 and AmpereOne and SPEC CPU INT 2017 on AmpereOne and Apple M5. Pixel SYNC mode incurs high overhead in some benchmarks. Certain ASYNC benchmarks also incurs high overheads, contradicting general expectations. Ampere and Apple's SYNC mode also shows some overhead spikes, albeit smaller.}}
\setlength{\aboverulesep}{0.6ex} 
\setlength{\belowrulesep}{0.9ex} 
\label{fig:glibc}

\end{figure*}

\subsection{Apple's MTE implementation}

Apple added MTE support under the name Memory Integrity Enforcement (MIE) on the M5 chip (Apple Macbook Pro M5) and A19 chip (iPhone 17)~\cite{apple-mie}. Both chips come with performance and efficiency cores, and like the Ampere implementation support only MTE's SYNC mode. Since these chips were announced shortly before this paper's submission, we only discuss our preliminary analysis of the M5's performance core, and do not analyze micro-architectural details.

\section{MTE for Probabilistic Memory Safety}
\label{sec:perf}

We start by exploring the overhead of MTE for probabilistic heap memory safety with SPEC (\S\ref{subsec:spec}) and server workloads (\S\ref{subsec:server}). We then explore the microarchitectural sources of these overheads with targeted microbenchmarks (\S\ref{subsec:mte-slowdowns}).

\subsection{Experimental setup}\label{e-setup}
\kwb{Processor/Memory:} Pixel 8 (Tensor-G3/12GB), Pixel (Pro) 9 (Tensor-G4/16GB), AmpereOne (A192-32X/512GB), MacBook Pro M5 (M5/32GB). \kwb{Operating systems:} Pixel 8-Android 14~(AP2A.240605.024), Pixel 9-Android 15~(AP4A.241205.013), AmpereOne-Fedora 42 / Linux kernel 6.14., MacBook Pro M5-MacOS Tahoe 26.2 (25C56)
\kwb{Build:} For each benchmark we use the same binary across all
devices except Apple M5. Binaries are compiled with Clang 18 and statically linked with glibc-2.36. MTE is enabled in GLIBC's heap allocator via tunables (\texttt{glibc.mem.tagging}) and run on \texttt{debootstrap}~\cite{debootstrap} environment on Pixel. Apple requires its own binaries, and we used a modified Mimalloc~\cite{mimalloc} allocator to tag every allocation\footnote{The default MTE allocator for MacOS is ``Xzone Malloc''~\cite{apple-xzone}. We do not use Xzone for our benchmarks as it employs opaque tagging policies and its source code was not publicly available until recently.}.

\kwb{Frequency Pinning:} Every task is pinned to a specific core. Pixel 8 and 9—80\% of maximum allowed frequency to keep down benchmarking times while still providing thermal stability. AmpereOne--we pin the frequency to 2.6GHz for all benchmarks. Apple M5 does not support core/frequency pinning. All MTE modes (SYNC, ASYNC, ASYMM) are evaluated on Pixel, while we evaluate only SYNC on AmpereOne and Apple M5 as they don't support other modes(\S\ref{subsec:ampere-mte}).


\subsection{Probabilistic memory safety on SPEC CPU}
\label{subsec:spec}

\Cref{fig:glibc} shows the MTE overhead for heap memory safety on SPEC CPU 2006~\cite{spec2006} for the Pixel 8 and Ampere cores. Pixel 9's performance is similar to Pixel 8, and is available in Appendix \ref{sec:extra-figures}\ifextended{(Figure~\ref{fig:spec-analogs-pixel9})}. We run SPEC CPU 2017 on AmpereOne and Apple M5 as only these devices have sufficient memory to run this benchmark.
The coefficient of variation is typically under 3\%. A few programs like \texttt{libquantum} or \texttt{mcf} can have variation up to 5\%, so we increase the iterations for these programs to 10 iterations.

\para{Interpretation}
We observe substantial variation in performance degradation across benchmarks and cores.  Pixel's in-order Little cores show the smallest variation across workloads, with modest slowdowns of 20\% at most.

Pixel's out-of-order Big and Performance cores exhibit more variation. The Performance core in particular has up to a \glibcsynchmmerx slowdown on \texttt{456.hmmer} and \glibcsynchrefx on \texttt{464.h264ref}. A similar $5\times$ slowdown was observed on the SPARC M7 ADI's implementation of memory tagging~\cite{kostya-tagging}, but was not analyzed/explained. 
Briefly, these slowdowns are attributable to store operations in the Performance core not being allowed to run speculatively.
We explore this in detail in \S\ref{sec:serial_store}. 

AmpereOne's SYNC implementation broadly shows low overheads on CPU 2006 with a geomean of \amperespecgeomean, with a notable jump for \texttt{456.hmmer} of \amperehmmer.
Briefly, we are able to attribute this overhead to an implementation gap in its store-to-load forwarding and find that the geomean of \amperespecgeomean would significantly reduce if this is fixed.
We explore this in more detail in \S\ref{sec:ampere-store-to-load}.

Despite MTE ASYNC being the suggested option for production use, it exhibits large overheads that are comparable to those of MTE SYNC when running \texttt{403.gcc}, \texttt{471.omnetpp}, and \texttt{483.xalanc} on the Big core. 
Briefly, we are able to attribute the overhead of MTE ASYNC in these three benchmarks to pipeline backend stalls—a behavior that is seen when the tag-checking mechanism is saturated by a mix of random and sequential loads. We explore this in detail in \S\ref{sec:async-big-core-analysis}.

SPEC CPU 2017 single-core results (\texttt{refspeed}) on AmpereOne and Apple M5 show low overheads---geomean of \amperespecseventeengeomean and \applespecseventeengeomean respectively.
Multi-core benchmarks (\texttt{refrate}) are similar, and are thus omitted from \Cref{fig:glibc} with one notable outlier---when \texttt{502.gcc\_r} is run on one of SPEC CPU 2017's multi-core inputs (input 5), it slows down by $1.25\times$ on AmpereOne and by $1.29\times$ on Apple M5. We attribute this to overheads of assigning tags in large allocations. We explore this in \S\ref{sec:selective-tagging}.

\subsection{Server Workloads}
\label{subsec:server}

We test MTE's probabilistic heap-memory safety performance with four popular server-side programs: RocksDB 1.7.0, Nginx 3.0.1, PostgreSQL 1.15.0, and Memcached 1.2.0~\cite{rocksdb, nginx, memcached, postgresql} using benchmark configurations from Phoronix's \texttt{pts/server} benchmark suite version 10.8.4.\footnote{This version of Phoronix's benchmarks had an error with inconsistent PostgreSQL configuration in its scripts. We fix this error in our tests.}

In addition to using \texttt{glibc}'s \sys{malloc}, some of these benchmarks also directly allocate memory with \texttt{mmap} or use alternative allocators. To ensure all heap data is tagged, we use \texttt{strace}\cite{strace} to identify these cases, and modified the source to add the \texttt{PROT\_MTE} flag to all allocators. Error bars in all server benchmark figures represent 95\% confidence interval. Additional details on setup are included in Appendix~\ref{appendix:server}.

\begin{figure}[t]
    \centering
    \includegraphics[width=0.9\linewidth]{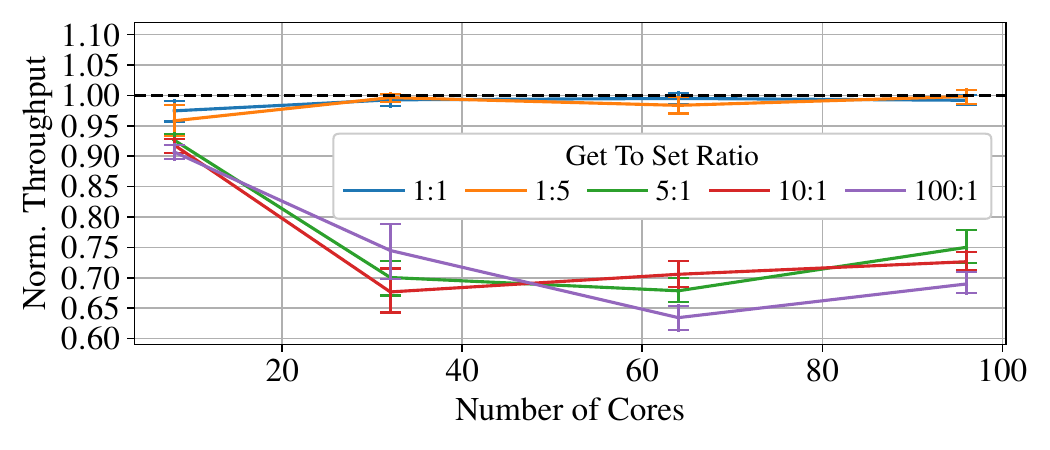}
    \caption{\textbf{MTE-enabled Memcached performance on the AmpereOne.} Memcached shows slowdowns of up to \amperememcachelargedropmax; However, we found this is due to performance problems in the Linux kernel's MTE support, rather than hardware limitations. }
    \label{fig:ampere-memcached}
\end{figure}

\para{Initial Evaluation}
We start by evaluating overheads of applications where servers and workload generators run on the same machine---analogous to deployments where scalable applications made of multiple services runs these services on shared physical hardware.
This setup also eliminates network noise and isolates CPU and memory performance highlighting overheads of MTE hardware.

We observe that overheads of RocksDB, Nginx, and PostgreSQL are similar to the overheads in application benchmarks like SPEC CPU 2006 (\S\ref{subsec:spec})---typically under \ampererocksdbmaxdrop, and negligible in some cases \ifmain{(figure available in Appendix~\ref{sec:extra-figures})}\ifextended{(Appendix~\Cref{fig:ampere-server-bench-other})}.

In contrast, Memcached showed a larger performance drops up to \amperememcachelargedropmax when running on many cores. In particular, these occur when the Set-to-Get ratio, i.e., the ratio of key-value writes to reads, skews heavily toward reads (Figure~\ref{fig:ampere-memcached}).
Further analysis showed that Memcached's overheads persisted even when MTE was never used by the application, but simply enabled at the kernel-level for the benchmarked process.

\begin{figure}[t!]
    \centering
    \begin{subfigure}{\linewidth}
        \centering
        \includegraphics[width=0.9\linewidth]{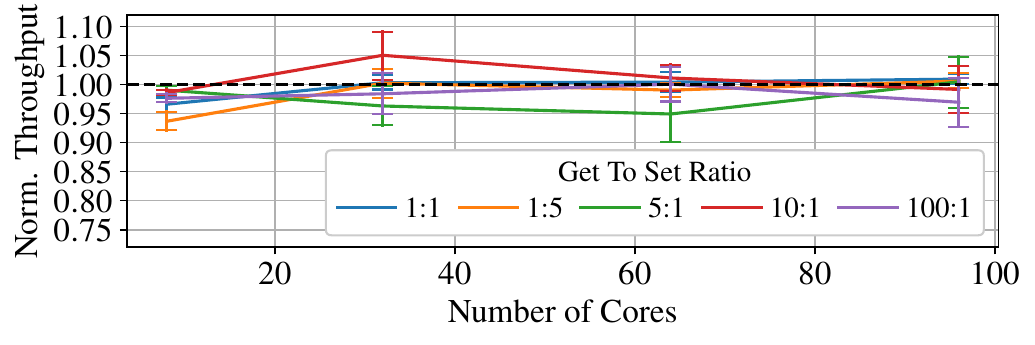}
        \caption{Memcached: querying key-value pairs with diff. get-to-set ratios.}
        \label{fig:ampere-memcached-fix}
    \end{subfigure}
    \begin{subfigure}{\linewidth}
        \centering
        \includegraphics[width=0.9\linewidth]{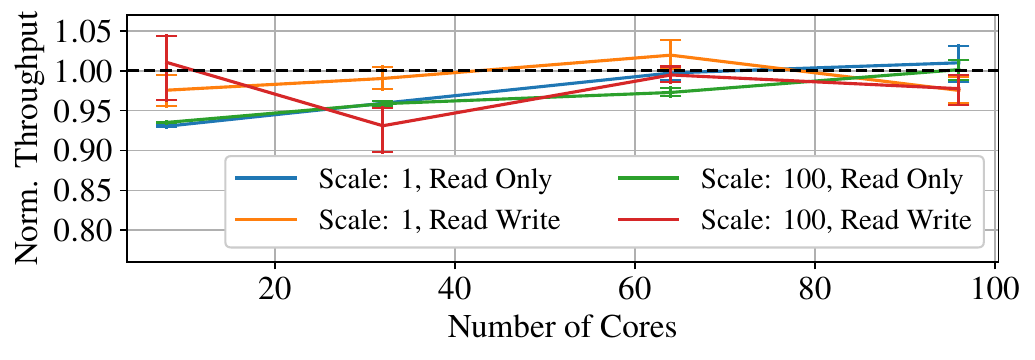}
        \caption{PostgreSQL: SQL queries on databases of different size.}
        \label{fig:ampere-postgres-fix}
    \end{subfigure}
    \begin{subfigure}{\linewidth}
        \centering
        \includegraphics[width=0.9\linewidth]{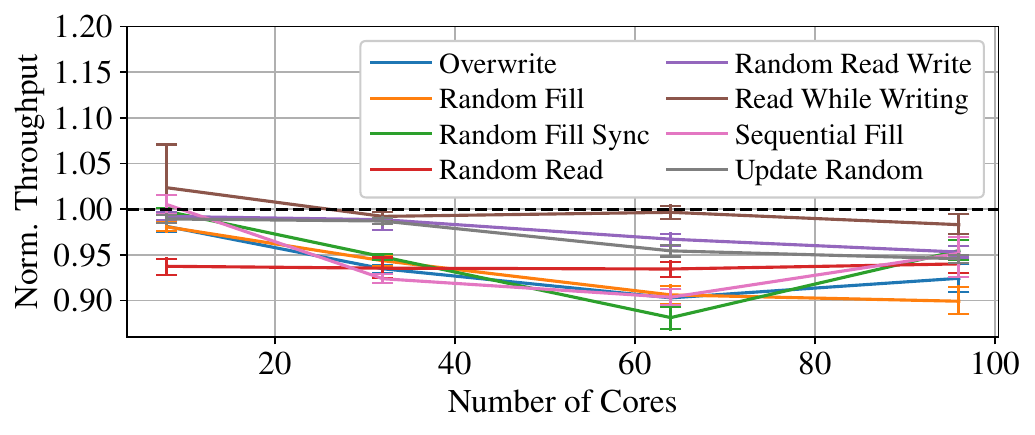}
        \caption{RocksDB: tests on different I/O patterns in database operations.}
        \label{fig:ampere-rocksdb-fix}
    \end{subfigure}
    \begin{subfigure}{\linewidth}
        \centering
        \includegraphics[width=0.9\linewidth]{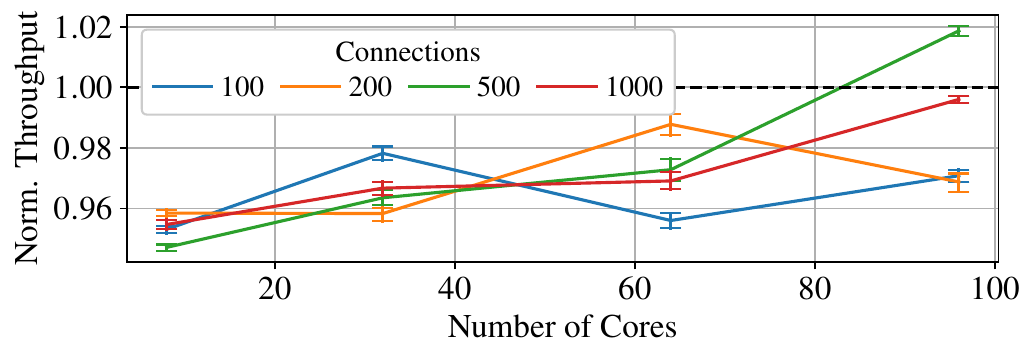}
        \caption{Nginx: tests on serving static files.}
        \label{fig:ampere-nginx-fix}
    \end{subfigure}
    \caption{\textbf{MTE server workload performance on AmpereOne after our kernel patch.} MTE overhead are generally under \amperefixmax. Most importantly, our patch successfully addresses the Memcached regression shown in \Cref{fig:ampere-memcached}, improving the worst case \amperememcachelargedropmax to about \amperefixmemcached overhead.}
    \label{fig:ampere-server-fix}
\end{figure}

\para{Unnecessary Tag Checks in Linux}
We narrowed the source of overhead to unnecessary tag checks in the Linux kernel that occur for two reasons.
First, the kernel marks all memory in its address space as "taggable" to preserve tag metadata on any user pages mapped in the kernel address space.
Second, the kernel ensures that accesses to kernel pages do not check tags---done by configuring three registers: the Tag Check Fault 0 (\texttt{TCF0}) and Tag Check Fault register (\texttt{TCF}) registers which toggles tag checks on user and kernel mappings respectively, and the Tag Check Override register (\texttt{TCO}) which disables checks on both accesses. 

The kernel code incorrectly assumes that any configuration that disables tag check faults via these registers also disables tag checks and thus avoid their overheads; however, ARM MTE's specification is ambiguous about this.
While Pixels disable tag checks altogether avoiding overheads, AmpereOne suppresses the faults and thus continue to incur tag check overheads when accessing the untagged kernel memory.

To resolve this, we initially developed a fix that configures these registers differently to avoid overheads (which we rely on for the rest of this section). However, we note that discussions with kernel reviewers~\cite{lkml-mte-patch} yielded a simpler solution that relies on a different configuration register---\texttt{TCMA1}---that explicitly disables tag checks for kernel memory accesses avoiding overheads.

\para{Evaluation with the kernel fix}
We demonstrate the efficacy of our initial fix in \Cref{fig:ampere-server-fix}. The patch eliminates the performance regression in Memcached, reducing the relative overhead from nearly \amperememcachelargedropmax (as seen in \Cref{fig:ampere-memcached}) to under \amperefixmemcached. Also, our patch improves the overall performance of other server apps, showing overheads under \amperefixmax. We have submitted this patch to the Linux Kernel upstream~\cite{lkml-mte-patch}.

\para{Multi-server setup}
To estimate the impact of MTE in a cloud deployment scenario, we additionally conduct multi-server tests using a remote load generator.
In this multi-server setup, we observe no statistically significant performance difference between runs with or without MTE for Memcached and Nginx. This result is consistent with numbers reported by Ampere's~\cite{ampere-arxiv-v2}\footnote{Ampere was motivated to conduct and release these benchmark numbers after we shared a draft of our paper with them (See Section~VIIB in \cite{ampere-arxiv-v2}). }. Applying our patch yields no changes to the overall throughput in the multi-server setup as well.

\subsection{{\textmu}-architectural causes of MTE slowdowns}
\label{subsec:mte-slowdowns}

We investigate the worst-case slowdowns in Pixel's Performance and Big cores and the AmpereOne core with microbenchmarks and analyze microarchitectural causes.

\subsubsection{MTE SYNC slowdown in Pixel's Perf. Core}
\label{sec:serial_store}

The Performance cores in both Pixel 8 and Pixel 9 (we report for Pixel 8 X3 core below) exhibit the highest overall overheads when enforcing probabilistic memory safety for \texttt{456.hmmer} (\glibcsynchmmerx) and \texttt{464.h264ref} (\glibcsynchrefx), as shown in ~\Cref{fig:glibc}.
Examining the assembly, we find that both benchmarks have a tight loop that includes store operations.

\para{Microbenchmark results}
We construct a simple, tight-loop, store-heavy microbenchmark that performs a store to a fixed address. This simple program exhibits a $7.4\times$ slowdown---similar to the slowdown of \texttt{456.hmmer}. We use this microbenchmark to explain the microarchitecture behavior.

\para{Interpretation}
We attribute the large slowdown to an implementation choice in the Performance core outlined in Section 4.15 of its optimization guide~\cite{x3-optimization-guide}—speculative stores are disallowed for MTE SYNC. If there are two consecutive stores on tagged memory, the second store instruction can only be executed after the first store's tag check has finished. The implication is that each store acts as a memory barrier (fence).

We confirm this by inserting a memory-store barrier (\texttt{DMB ST} instruction) after the store instructions within the loop and observing that: (1) the modified microbenchmark (with MTE disabled) slows down by $6.2\times$, and (2) the modified microbenchmark does not slow down more than the original with MTE SYNC. Hence, the MTE SYNC implementation in this core indeed acts as a store memory barrier.

We note that the SPEC CPU benchmarks and our microbenchmark do not show significant overheads on the Big core (A715), indicating that the limitations on MTE SYNC stores are not fundamental and can potentially be fixed with additional implementation effort.

\subsubsection{MTE SYNC slowdown in the Ampere core}
\label{sec:ampere-store-to-load}

The AmpereOne CPU generally imposes lower overheads to enforce probabilistic heap safety; however, \texttt{456.hmmer} still incurs an overhead of \amperehmmer in \Cref{fig:glibc}.

To identify the root cause, we run this benchmark using \texttt{perf} with both MTE enabled and disabled. We found that enabling MTE resulted in significant decreases in \texttt{ld\_from\_st\_fwd}, indicating that the performance drop is related to the implementation of store-to-load forwarding.

To test this theory, we disabled store-to-load forwarding on the AmpereOne using an option in Ampere's BIOS and re-ran the full CPU 2006 suite with and without MTE. This time, we observed negligible overheads (under $1\%$) for \texttt{456.hmmer} when enabling MTE; in fact, the geomean of the entire suite fell from \amperespecgeomean to \amperespecnostlgeomean.

As part of other experiments in this paper (from \S\ref{sec:colorguard}), we identified an additional smaller benchmark that replicated this pattern with more exaggerated effects: a simple mathematical benchmark, \texttt{jacobi-2d} from the PolyBench/C numerical benchmark suite~\cite{polybench}, displayed overheads of \amperejacobinative when MTE was enabled\footnote{When compiled natively as done here, \texttt{jacobi-2d} has higher overheads than when compiled with sandboxing compilers in \S\ref{sec:colorguard}}. It also showed a drop in \texttt{ld\_from\_st\_fwd} in \texttt{perf}, and its overhead was also mitigated when store-to-load forwarding was disabled.

\para{Interpretation}
Our testing points to a limitation in the store-to-load forwarding logic in the AmpereOne when MTE is enabled. However, from what we know of MTE implementations (\S\ref{sec:mte-impls}), there is no reason why this points to a fundamental challenge. Logically, store-to-load forwarding circuits in a CPU can be made MTE-safe by modifying store buffers to track the MTE tag of the stores and checking that any store forwarded has the same tag as the load it is fulfilling.

Thus, it must be the case that  AmpereOne either does not implement the logic to carry tags in the store buffer, or the optimization works inconsistently when MTE is enabled. Given that we still observe some successful store-to-load operations in \texttt{perf} when MTE is enabled, we believe it is the latter.

We reached out to Ampere to check our interpretation. Ampere confirmed that the AmpereOne CPU has an MTE-related performance issue where MTE tags sometimes interfere with store-to-load forwarding, which caused the overheads we observed in \texttt{456.hmmer} and \texttt{jacobi-2d}. More interestingly, they confirmed that this was already resolved in their next-generation implementation and will recover the lost performance, validating our hypothesis that this is not a fundamental issue for MTE implementations.

\subsubsection{MTE-ASYNC/SYNC slowdown in Pixel's Big Core}
\label{sec:async-big-core-analysis}
Unlike on Pixel's Performance core, several benchmarks such as \texttt{403.gcc} (\glibcasyncgccbig), \texttt{471.omnetpp} (\glibcasyncomnetbig), and \texttt{483.xalancbmk} (\glibcasyncxalancbig) show significant overhead under both \emph{ASYNC and SYNC} modes on Pixel's Big Core (\Cref{fig:glibc}).
This is particularly troubling, as MTE ASYNC's goal is to minimize overheads for production use.

We re-ran these workloads using \texttt{perf} and found that the overhead stems from increased back-end stalls (consistent with prior observations from Gorter et al.~\cite{gorter2024sticky}). Further analysis revealed that these stalls are caused by code patterns that involve a mix of randomized and sequential memory accesses\footnote{\texttt{403.gcc}'s \textsf{reg\_is\_remote\_constant\_p} function exhibits this pattern.}. To better understand these hazards and why they cause slow downs, we construct a microbenchmark with a parameterizable version of this pattern.

\para{Microbenchmark construction}
Our microbenchmark consists of a link-list of $L$ nodes, where each node has a pointer to a byte-array of size $A$, resulting in a total of $L{\times}A$ bytes. 
The microbenchmark iterates through the linked-list from head to tail. For each visited node, it strides through the node's array and sums up every $S$-th element. 
The pseudocode for this microbenchmark is shown in the Appendix \Cref{alg:ll-bench}.

Running this microbenchmark with different values of $(L, A, S)$ allows us to test different aspects of the microarchitecture. For example, we can control the ratios of randomized $(L)$ vs. sequential $(A/S)$ loads—this controls metrics such as cache hits/misses, in-flight/pending memory access instructions, etc. Changing $A$ and $S$ also allows us to test the effects of microarchitectural components like the stride prefetcher.

\para{Microbenchmark results}
We run this benchmark 10 times after 10 warm-up iterations on the Pixel Big core with MTE ASYNC, and report overheads normalized to when MTE is disabled. 
We repeat the benchmark for a range of values of $(L, A, S)$ and show an interesting sample of these results in \Cref{fig:glibc-big-core}---namely for a range of values of $L$ and $A$, when $S$ is set to 4B and 128B. 

We observe that: 
(1) the overhead of MTE ASYNC becomes noticeable when $A\times L \geq 8$MiB---the size of the Pixel's last level cache,\footnote{We reverse-engineered the size of system last level cache, which aligns with prior industry tech reports~\cite{GoogleP8Explained}} and
(2) the maximum overhead is sensitive to the stride $S$, ranging from $1.5\times$ at $S=4$ to $3.8\times$ at $S=128$.
As a sanity check, we used \texttt{perf} to ensure that the slowdown has the same root cause (increased back-end stalls) as the three slow SPEC CPU benchmarks.

\begin{figure}
    \centering
    \includegraphics[width=\linewidth]{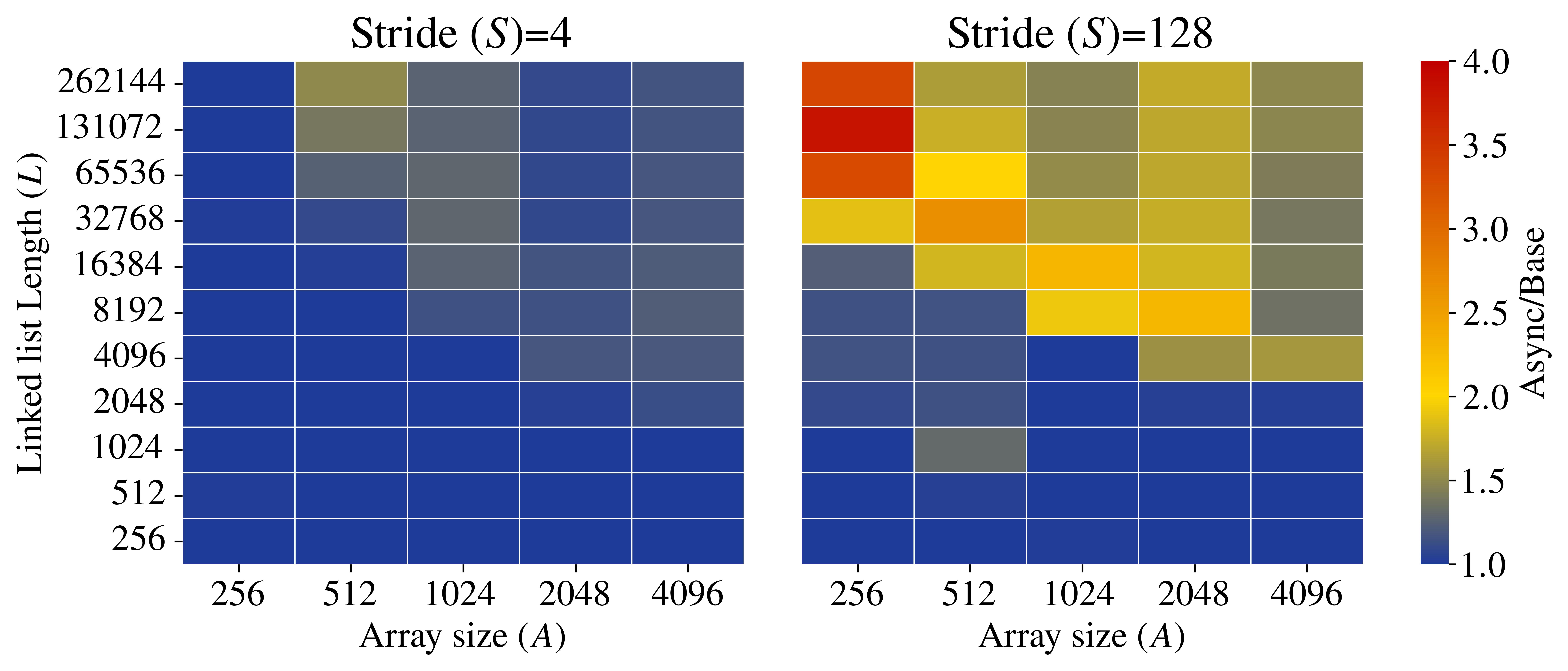}
\caption{\kwb{Uncovering structural hazards due to MTE ASYNC.}\kw{
Using a stride access pattern within small arrays stored in a linked list generates 
cache misses which stress the tag-load/tag-check operations in the Pixel Big
Cores data path. This reveals structural hazards that bottleneck loads, slowdowns
can be seen in the heatmap---lighter is slower.}}
\label{fig:glibc-big-core}
\end{figure}

\para{Interpretation}
We attribute the slowdowns to structural hazards---bottlenecks in the Pixel Big core’s concurrent MTE tag-checking and memory access hardware---and find that the overheads of the three slow SPEC CPU benchmarks are approximated by a combination of overheads at different strides.

Such hazards are possible as MTE support fundamentally requires hardware to perform and track two steps in each memory operation---the tag check and the actual memory access. This means that when MTE is enabled, the ability to execute a memory instruction without stalling, is contingent on the CPU being able to issue a tag check operation (even in MTE's ASYNC mode).

While we cannot identify the exact microarchitectural structure that saturates under tag checking without white-box access to ARM's MTE design, our parameterizable microbenchmark sheds light on the phenomenon.

Concretely, when the size of the link-list exceeds the last-level-cache size ($A \times L \geq 8$MiB), both the random pointer dereference (one per array) and the subsequent $A/S$ striding accesses to the array likely cause cache misses requiring both data and tag fetches from memory. The resulting accesses fill the microarchitectural structures that track in-flight accesses, causing stalls and degrading performance (in both SYNC and ASYNC modes). %
Specifically, we observe two scenarios:

\textit{When $S \leq 64$:} We see that the maximum overhead is $\approx 1.5 \times$ and appears only when $A$ is sufficiently large ($\approx 512 B$). Our interpretation is that when both $A$ and $S$ are small, the $A/S$ strided accesses are largely covered by cache hits and next-line/stride prefetchers and thus the overhead vanishes. 

\textit{When $S \geq 128$:} We see the maximum overhead jumps to $3.0 \times$, especially when $A$ is small (256).  Our hypothesis is that at this large stride, there is insufficient chance for prefetching and there are no cache hits. As a result, more pressure is placed on in-flight tracking hardware and prefetchers cannot compensate by working ahead of the demand miss stream.

We believe such structural hazards can be fixed for future implementations. As one data point, we observe that both SPEC and our microbenchmarks don't show significant overheads on Pixel's Performance core in the MTE ASYNC mode. This indicates that such overheads are not fundamental.

\subsubsection{Tagging cost on AmpereOne and Apple M5}
\label{sec:selective-tagging}

\texttt{502.gcc\_r} in \Cref{fig:glibc} shows relative high overhead on AmpereOne and Apple M5. We observe this often allocates large memory regions, thus we hypothesized that this slowdown is due to the cost of instructions for tagging these regions (e.g., \texttt{stg}).

To validate our hypothesis, we modified the allocator to selectively tag only allocations smaller than 32KiB. Indeed, several production allocators employ similar policies~\cite{partap2022memory, scudo, apple-xzone}. For large allocations, these allocators instead rely on guard pages (unmapped pages after an allocation) to prevent some buffer overflows. 

The performance difference is marginal for most workloads in SPEC, showing 0-2\% improvement in the geomean across the tested cores.
However, specific benchmarks show notable improvements for some micro-architectures. For \texttt{403.gcc} (SPEC 2006) overhead dropped from  \amperegccselectivebefore to \amperegccselectiveafter on AmpereOne and from \peightlittlegccselectivebefore to \peightlittlegccselectiveafter on Pixel 8 Little Core. 
Similarly, SPEC 2017's \texttt{502.gcc\_r} benchmark with input 5 (discussed in \S\ref{subsec:spec}) decreases from $1.25\times$ to $1.17\times$ on AmpereOne and from $1.29\times$ to $1.08\times$ on Apple M5.

\para{Interpretation} While the cost of instructions like \texttt{stg} are not typically bottlenecks, their overhead is noticeable in specific workloads and microarchitectures for large allocations.

\section{Additional Applications of MTE}
\label{sec:flex}

We analyze MTE's performance in a range of different applications beyond memory-safety to explore its flexibility. In some cases, it provides advantages (our positive results). In other cases, it is equivalent or inferior to existing hardware or software mechanisms (mixed to negative results). Unless noted otherwise, our experimental setup is similar to \S\ref{sec:perf}.

\subsection{Positive Results: BufLock \& Mem. Tracing}
MTE can be used as a fast substitute for copying to prevent time-of-check time-of-use attacks. It also offers a memory tracing mechanism with better performance than page protections and better scalability than hardware watchpoints.

\subsubsection{TOCTOU protections via revocation (BufLock)}
\label{sec:toctou}

Prior work has observed that MTE offers efficient time-of-check time-of-use (TOCTOU) mitigations by allowing revocation of buffers from untrusted code~\cite{chen2024limitations}. We demonstrate this with an example in the Firefox browser. Firefox runs its XML parser in an in-process (SFI) sandbox~\cite{wahbe1993efficient, rlbox}. Concretely, this means that the XML parser runs in a dedicated memory pool---the sandbox memory---and cannot access other memory even in the presence of memory-safety bugs.

For security, Firefox must copy any data produced by the XML parser out of the sandbox memory prior to use; this ensures a compromised XML parser cannot unexpectedly modify parsed data as a means to attack Firefox, i.e., a TOCTOU attack\footnote{This is analogous to OS kernels copying system call parameters from user memory to kernel memory prior to using the data.}. However, despite requiring data copies to preserve security, copying all data generated by the XML parser results in a 10\% slowdown---something too slow to deploy~\cite{firefox-libexpat-sbx-perf}.

The TOCTOU vulnerability here is caused by code re-entrancy (rather than threading). Concretely, Firefox repeatedly makes calls to the sandboxed XML parsing library to parse chunks of an XML file as it comes in over the network, storing the parsed data in sandbox memory. Firefox checks the parsed data and assumes that the already parsed XML data will not change during subsequent calls to the XML parsing library; however, a compromised or malicious XML parser can violate this expectation. This would result in a TOCTOU bug in Firefox when the data is next used.

\para{BufLock}
MTE offers an alternative to slow data copies~\cite{chen2024limitations, xia2019cherivoke}. If we assume the XML parser can't use any tagged pointers, we can prevent it from conducting TOCTOU attacks just by tagging these data structures with any tag (BufLock). This is, in effect, a poor man's capability revocation mechanism.

The remaining question: how do we ensure the sandboxed XML parser doesn't use tagged pointers (i.e., pointers with non-zero tag bits)? Fortunately, SFI mechanisms such as WebAssembly~\cite{wasm} used by Firefox~\cite{rlbox} guarantee this automatically. SFI works by employing a custom compiler that inserts runtime checks to ensure all dereferenced pointers have a known prefix~\cite{wahbe1993efficient}. This also guarantees that the tag bits of pointers are zero. Additionally, SFI compilers are, by default, configured to disallow any instructions that attackers may find useful---including instructions like \texttt{stg} which can change tags on memory, or system calls that disable MTE.

\begin{figure}[]
\centering
\scriptsize
\setlength{\aboverulesep}{0pt} 
\setlength{\belowrulesep}{0pt}
\begin{tabular}{l|cll} \toprule
Cores & \makecell{Insecure\\ (sec)} & Data copies & BufLock \\ \midrule
Pixel8 (Little) & 1.19  & 1.28 (\bufcopyPl)     & \cellcolor{green!25}1.20 (\buflockPl)     \\ \midrule  
Pixel8 (Big)    & 0.31  & 0.35 (\bufcopyPb)     & \cellcolor{green!25}0.32 (\buflockPb)     \\ \midrule  
Pixel8 (Perf)   & 0.17  & 0.19 (\bufcopyPx)     & \cellcolor{red!25}0.24 (\buflockPx)       \\ \midrule  
Pixel9 (Little) & 0.86  & 0.92 (\bufcopyPIXl)   & \cellcolor{green!25}0.89 (\buflockPIXl)   \\ \midrule  
Pixel9 (Big)    & 0.23  & 0.25 (\bufcopyPIXb)   & \cellcolor{green!25}0.25 (\buflockPIXb)   \\ \midrule  
Pixel9 (Perf)   & 0.13  & 0.14 (\bufcopyPIXx)   & \cellcolor{red!25}0.18 (\buflockPIXx)     \\ \midrule  
AmpereOne       & 0.23  & 0.25 (\bufcopyampere) & \cellcolor{green!25}0.24 (\buflockampere) \\ \midrule  
\end{tabular}
\caption{\textbf{Replacing Copying with MTE for Fast TOCTOU Protection.} \textit{Replacing memcpy with re-tagging (BufLock) can often reduce the cost of TOCTOU protection when sharing data across isolation boundaries. Cases where BufLock reduces these overheads for protected buffers from an XML parser (sandboxed libexpat) in Firefox are shown in green.}}
\label{tbl:buflock}
\setlength{\aboverulesep}{0.6ex} 
\setlength{\belowrulesep}{0.9ex}
\end{figure}

\para{Experimental setup}
While our experimental setup is broadly similar to \S\ref{sec:perf}, Firefox is compiled as a native Android application using the native \texttt{bionic} libc (rather than a Debian application that is run on \texttt{debootstrap} with \texttt{glibc}). Additionally, to avoid noise, we modified Firefox to pin only its XML parsing thread to an isolated core with a pinned frequency; all other processes/threads run on different cores.

\para{Evaluation}
We used BufLock to "lock" the data structures produced by Firefox's sandboxed XML parser and thus prevent TOCTOU attacks. We use the same benchmark used by Firefox engineers to test the performance of the XML parser~\cite{firefox-libexpat-sbx-perf}: measuring the time taken to parse a large SVG image (represented in XML) from Google Docs, tiled 10 times vertically. After ignoring 10 runs, we report the median time to render the SVG image 100 times. Figure~\ref{tbl:buflock} shows the performance of three Firefox builds---an insecure version that neither copies data nor uses BufLock to prevent TOCTOU, one that copies data, and one that uses BufLock.

\para{Interpretation}
Copying memory and managing extra allocations have visible CPU performance overheads (between \buflockPIXl and \buflockPx with the coefficient of variation under 3\%); this is consistent with the $1.10\times$ overhead reported by Mozilla engineers~\cite{firefox-libexpat-sbx-perf}.

BufLock significantly reduces these overheads on the AmpereOne core as well as the Pixel Big and Little cores, which makes shipping protections for this vulnerability much more practical. However, overhead on the Pixel's Performance core goes up significantly, caused by the limitation of Pixel's Performance core identified in \S\ref{sec:serial_store}. However, if this bug in the Performance core is fixed, BufLock would offer a meaningfully faster way to secure this operation.

To offer additional insight into why BufLock's tagging approach is faster than copying, we implemented a microbenchmark on these operations in Appendix\ifmain{~\ref{appendix:extra-microbench}}\ifextended{~\ref{sec:stg}}. We observe that using the optimal mix of tag instructions (\texttt{dc gva} to tag entire cache lines, \texttt{st2g} and \texttt{stg} to tag two and one granules), results in better performance than copying data.

\para{Multi-threading}
While BufLock works well in this single-threaded use case, its costs in a multi-threaded program could be more expensive given MTE's current design (\S\ref{sec:wishlist}).

\subsubsection{Memory Tracing}
\label{sec:trace} 

Efficiently tracing data access can support interactive debugging~\cite{cui2018rept}, data profiling for cache optimization~\cite{brais2020survey}, dynamic program analysis~\cite{gosain2015survey, greathouse2012case}, and intrusion detection~\cite{suneja2015exploring}. 
Unfortunately, software binary-translation approaches such as PIN~\cite{pin} and DynamoRio~\cite{dr-paper} are complex and often expensive; hardware supports only a limited number of watchpoints (usually 4 to 8); and using page permissions for tracing is expensive due to the high costs of changing page permissions and the problem of spurious traps due to false sharing.

MTE offers the potential to enable an unlimited number of efficient hardware data watchpoints, and other work has already put this to good use for efficient online data race detection~\cite{shastri2024hmtrace}. To explore this capability, we developed two MTE-powered tracing tools: "MTE-tracer", which leverages MTE to trace accesses to data structures from user space, and "MTE-kernel-tracer", which is an additional optimization that pairs the MTE-tracer approach with a custom kernel module for better performance. We also built "Page-tracer", a tracer leveraging page permissions; we use Page-tracer and DynamoRIO~\cite{dr-paper} as baselines in our evaluation.

\para{Implementing MTE-tracer and Page-tracer} MTE-tracer's basic approach is to simply tag any memory/data structures we want to trace with a chosen tag. Thus, any accesses to these data structures will result in a segmentation fault with an extra flag indicating an MTE tag mismatch.

When a fault occurs, the tracer executes a routine we call \texttt{log-step-and-resume}; it logs the access, temporarily untags the data, re-executes the faulting instruction, retags the data, and then resumes normal execution. While one can leverage single-stepping like a standard debugger to perform these steps, the overhead of redundant signals/exceptions and following context switches is prohibitive. To avoid this, MTE-tracer dynamically generates a code snippet for the \texttt{log-step-and-resume} sequence and executes that code at every fault instead of every single step.

We also implement Page-tracer, which uses page permissions rather than MTE to provide another baseline (in addition to DynamoRIO). Page-tracer uses essentially the same approach; however, it must (1) use expensive \texttt{mprotect} system calls in place of retagging data and (2) check that the segmentation fault was not due to false sharing (i.e., the bytes on the page being accessed are not those being traced).

\para{Implementing MTE-kernel-tracer}
The control-flow path of MTE-tracer is unnecessarily complex and expensive as using signal handling for event delivery requires four user-kernel context switches for every event. MTE-kernel-tracer adds a kernel module to optimize these transitions away by directly logging events in memory in the kernel, which user space can periodically drain. The module works by using \texttt{kprobes} to bypass the Linux kernel's normal MTE fault handling and runs \texttt{log-step-and-resume} instead. After this, execution of the process is resumed. This approach only requires two user-kernel context switches for every event.

\begin{figure}[t]
    \centering
    \begin{subfigure}[b]{0.48\linewidth}
        \centering
        \includegraphics[width=\linewidth]{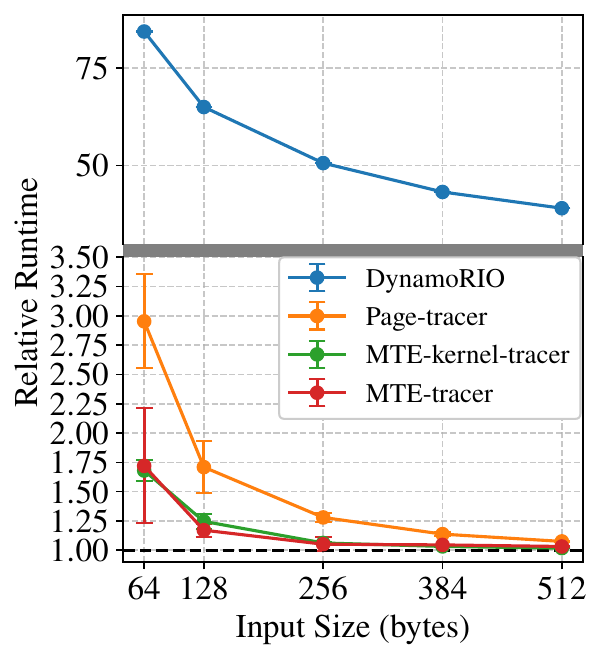}
        \caption{RSA Sign}
        \label{fig:memtrace-sign-pixel-perf}
    \end{subfigure}
    \hfill
    \begin{subfigure}[b]{0.48\linewidth}
        \centering
        \includegraphics[width=\linewidth]{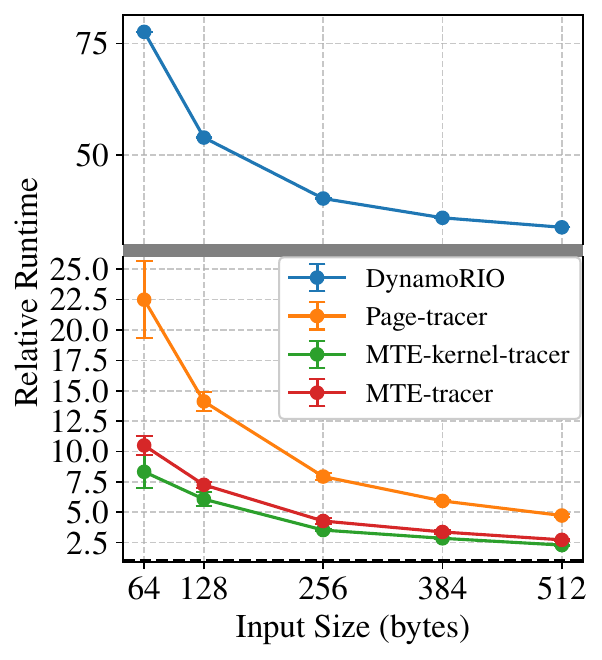}
        \caption{RSA Verify}
        \label{fig:memtrace-verify-pixel-perf}
    \end{subfigure}
    \caption{\kwb{MTE-based and Page-based data tracing overheads on Pixel 8 Performance Core:} \kw{We compare the performance of tracing key material through OpenSSL using different tracers; the time taken is normalized to the baseline performance with no tracing enabled. We see that the MTE-based tracer is orders of magnitude faster than DynamoRIO and is 2 to 3 times faster than the Page-tracer for small-sized inputs, and the gap decreases as the buffer size increases.}}
    \label{fig:memtrace-p8x}
\end{figure}

\para{Setting up DynamoRIO}
We configure DynamoRIO to instrument every memory access to check if the access is in an address range that is being traced. If so, the instrumentation logs the access and continues execution. 

\para{Benchmark setup}
We evaluate the overheads of memory tracing on OpenSSL using two of its supported performance ("OpenSSL-speed") benchmarks: performance of RSA key signing and RSA key verification. We measure the slowdown due to tracing memory accesses to a variable-sized input buffer---from 64 bytes up to 1024 bytes---containing a digital signature passed to the sign and verify operations. Since the benchmark accesses the buffer linearly, the size of the buffer is directly correlated to the number of traced memory accesses in this program. We add 10 warm-up runs prior to the benchmark to eliminate any one-time costs in benchmark setup.

OpenSSL normally runs these benchmarks for 10 seconds using a timer that issues a \texttt{SIGALRM} signal to terminate the benchmark.
However, as DynamoRIO support for \texttt{SIGALRM} is buggy~\cite{dynamorio-sigalrm}, we modify these benchmarks to run for a fixed amount of iterations instead---1000 for RSA Signing, 10000 for RSA Verification excluding warm-up runs.
 
\para{Minimizing unrelated costs in baselines}
We use DynamoRIO and PageTracer as baselines for evaluation. DynamoRIO's dynamic instrumentation is normally applied on the first run of the program, which could increase overheads on the first run; however, any costs associated with this are excluded because we use warm-up runs. PageTracer would normally encounter a large number of spurious traps due to false sharing; however, we place the input buffer being traced in its own dedicated page so we only consider the overheads due to the use of \texttt{mprotect} calls in the PageTracer.

\para{Evaluation}
Figure~\ref{fig:memtrace-p8x} shows the performance of different tracing approaches on the Pixel 8 performance core, along with 95\% confidence intervals. Page-tracer incurs throughput slowdowns of \tracemprotsignpixel and \tracemprotverifypixel for RSA sign and RSA verify for small buffers.
In contrast, MTE-tracer is much faster incurring slowdowns of \traceusersignpixel and \traceuserverifypixel, while MTE-kernel-tracer is faster still and incurs only \tracekernelsignpixel and \tracekernelverifypixel slowdowns. Performance of the other cores are similar, with minor variations such as the Ampere core being a bit faster; more data is available in Appendix\ifmain{~\ref{sec:extra-figures}}\ifextended{in Figure~\ref{fig:memtrace-all}}.

\subsection{Mixed Results: Sandboxing, CFI}
We investigate if ARM MTE can optimize sandbox enforcement~\cite{wahbe1993efficient} and control-flow integrity~\cite{abadi2009control}---properties typically enforced using compiler-inserted runtime checks. Broadly, we see that MTE is not ideal for sandboxing due to high and variable overheads, but may offer reasonable performance depending on the workload and hardware. MTE does not clearly provide meaningful performance benefits for CFI but can offer additional options for applications using JIT compilers.

\subsubsection{Enforcing/Optimizing Sandboxing schemes}
\label{sec:colorguard}

Prior work~\cite{chen2024limitations, segue-plas,segue-cg, cage} has proposed using MTE for enforcing sandboxing/SFI-style isolation. The idea behind these approaches is that a program component may be sandboxed if all of its memory accesses unconditionally use a fixed MTE tag. This guarantees that the isolated component can only access memory tagged with that fixed tag, isolating it from other pages. We use ColorGuard~\cite{segue-plas,segue-cg}'s implementation of this idea, as the other implementations mentioned above~\cite{chen2024limitations, cage} require tagging an application's entire private memory (rather than just the sandbox memory pages), meaning these incur much higher overheads due to MTE (See \S\ref{subsec:cage}).

ColorGuard is an optimization that lowers the virtual memory footprint of existing SFI tools like WebAssembly (Wasm)~\cite{wasm} and Native Client~\cite{nacl64}. These SFI tools enforce isolation by inserting runtime bitmasking instructions or bounds-checking instructions to ensure all memory addresses accessed by code remain in a specified memory region---\emph{the sandbox memory}~\cite{wahbe1993efficient, lefeuvre2025sok}. To optimize these runtime checks, SFI tools also place large regions of unmapped memory called guard regions (4 GiB in size in Wasm and 80 GiB in size in Native Client) after the sandbox memory. This allows SFI tools to elide runtime checks for any memory accesses that can statically be shown to remain within the guard regions.

ColorGuard observes that SFI tools can eliminate guard regions between adjacent sandboxes but continue to leverage the above guard-region optimizations if they ensure that adjacent sandboxes each use a different Memory Colors/Tag. In this case, adjacent sandboxes act like guard regions for each other. The original paper implemented Memory Colors leveraging the efficient page-coloring of Intel MPK~\cite{intel-manual}; they also suggested ARM MTE could also be used for ColorGuard but stopped short of testing this claim.

\para{Evaluation}
We implemented ColorGuard-MTE in wasm2c~\cite{wasm2c} Wasm compiler. In our implementation, as with ColorGuard-MPK, each sandbox is tagged with a single color that does not change during the course of its lifetime. The per-sandbox tag is stored as part of the "sandbox memory" address used for runtime checks; this guarantees that the sandbox is restricted to the specified tag.

We evaluated ColorGuard-MTE with PolyBench/C, a numerically heavy benchmark suite~\cite{polybench} that is frequently used to benchmark Wasm SFI compilers~\cite{wasm, ewasm, twine}. Each program in PolyBench/C runs 10 iterations, and we report the median of them with normalized minimum and maximum as error bars. Note that, in contrast to probabilistic memory safety, there is no runtime overhead from retagging; all overheads are due to the costs of tag checking.

\para{Interpretation}
The results of our evaluation are in Appendix \ref{sec:extra-figures}.
While the performance on Pixel's Little and Big core and the Ampere core is competitive for many workloads, the worst-case overheads of the Pixel Little, Big, Performance, and Ampere cores are \polybenchlittlemax, \polybenchbigmax, \polybenchxmax, and \polybenchamperemax respectively\ifextended{ (Figures~\ref{fig:polybench-pixel8} and \ref{fig:polybench-ampere})}.
This is noticeably higher than the typical single-digit overheads we expect from modern ARM SFI tools~\cite{lfi}.
This overhead stems from the bottlenecks discussed in Section~\ref{subsec:mte-slowdowns}; however, such slowdown is exacerbated because Wasm moves stack arrays onto the heap by design (similar to SafeStack~\cite{cpi}).

Consequently, this result is mixed. In general, the runtime overheads imposed by 
 MTE make it a poor replacement for ColorGuard's usual page-level tagging 
 mechanisms~\cite{segue-cg} that impose no runtime overhead. However, since no current ARM hardware ships with page-level tagging (POE~\cite{poe}), ColorGuard-MTE may still be a reasonable option.

\subsubsection{Enforcing/Optimizing Control-Flow Integrity} 
\label{sec:cfi}

Forward and backward edge control flow integrity (CFI~\cite{abadi2009control}) are popular techniques for exploit mitigation with both software~\cite{clang-cfi} and hardware support~\cite{pacstack,armpac,intelcet,riscv-cfi}. Prior work has suggested that tagging could be used to optimize CFI~\cite{hdfi,cctag}.

We evaluate if ARM MTE can be used to optimize CFI. We focus on two techniques: first, we optimize software-CFI by using tag checks to eliminate CFI's runtime checks; second, we tag code pointers to protect their integrity~\cite{cpi}. In general, we find that tagging with current MTE hardware does not offer benefits over existing software/hardware approaches.

\para{Backward-edge CFI}
We developed two forms of backward-edge CFI.
First, we introduce an enhanced shadow stack that isolates the shadow stack using MTE.
This is compelling, as current software CFI such as Clang's ShadowCallStack~\cite{clangSS}
relies only on randomization to protect the shadowstack, leaving it potentially vulnerable to attack.
Prior work~\cite{burow2019sok, cerberus} leverages MPK to protect the shadow stack on x86 platform, 
and MTE may be a drop-in replacement on the ARM platform.
Second, we tag return addresses on the stack to protects the integrity of return addresses in place. This is attractive as the return address does not have to be duplicated or moved from the stack frame; thus, \texttt{longjmp} in C, or bail-outs in JIT compilers~\cite{v8-deoptimization} are easier to support.

\para{Evaluation}
Our implementation of both schemes in LLVM is provided in  Appendix~\ref{appendix:cfi} and evaluated on SPEC CPU 2006.

We observe that while using MTE to isolate shadow stacks is fast, incurring only \tagssallcoresgeomeanmin to \tagssallcoresgeomeanmax across all cores (Appendix~\ref{appendix:cfi}\ifextended{, Figure~\ref{fig:tagss}}), this is comparable performance to existing approaches like PACStack~\cite{pacstack}—a technique for backward-edge CFI using code pointer signing and MACs.

We observe that using MTE to preserve the integrity of return pointers on the stack incurs prohibitively expensive overhead on the Pixel big and performance cores of \ralbiggeomean and \ralxgeomean respectively (even if they are reasonable on other cores) due to the cost of frequent retagging\ifextended{ (Figure~\ref{fig:ral})}.

\para{Forward edge CFI}
For forward-edge CFI, we explored replacing software CFI checks (e.g., on virtual tables and indirect jump tables) with MTE tag checks in the hopes of improving performance. Tagging also provides an alternative to placing landing-pad instructions for forward-edge CFI~\cite{abadi2009control} in code (e.g., inserting ARM64's \asm{bti} instruction at valid jump targets); applications can use tagging to store valid targets as vtables protected from modification by tags.

This is compelling, as ARM64 JIT compilers often place large program constants at small offsets from functions, i.e., in code pages, to enable efficient (PC-relative) addressing. Since attackers can often control these constants, this also opens the door to forging new \asm{bti} instructions\footnote{This is a known problem in V8~\cite{v8-cfi}, and RISC-V's recently introduced branch target instruction (\asm{lpad}), even supports an optional 20-bit immediate with a hash of the function signature to mitigate such attacks.}.

\para{Evaluation}
Our implementation of both schemes in LLVM is provided in Appendix~\ref{appendix:cfi} and evaluated on SPEC CPU 2006.
Unfortunately, we observe that MTE's performance penalties cancel out any speedups tagging offers to software-CFI, and the end result is pretty similar performance\ifextended{ (Appendix Figure~\ref{fig:fwd-cfis-clang})}. However, tagging-based CFI may still offer some value in JIT environments where its competitive performance with out-of-band landing pads offers benefits.
\section{Analyzing prior work on MTE performance}
\label{sec:prior-misconceptions}

A variety of prior work has attempted to characterize MTE performance. Some~\cite{hemate, safe-bpf, hakc, liljestrand2022color, sfitag, thesis-securing, wang-opportunistic, mtsan, zometag, capacity} used performance analogs---instruction snippets chosen to approximate the costs of MTE. While others~\cite{cage, gorter2024sticky, bastag} used MTE in the Google Pixel 8. Unfortunately, the performance analysis in these papers is often either incomplete or incorrect due to methodological limitations or implementation errors.

\subsection{Accuracy of MTE performance analogs}
\label{subsec:analogs}

In the absence of real hardware or cycle-accurate simulators with MTE support, emulating MTE's cost with analogs was the only option for a number of years. However, now that hardware is available, we can test the accuracy of these analogs and the conclusions reached using them.

We focus on two analogs for MTE's ASYNC mode introduced by HAKC~\cite{hakc} and SFITag~\cite{sfitag}, which were subsequently used in most other papers. The HAKC analog uses regular memory store instructions to simulate the costs of tagging instructions like \texttt{stg} and assumes tag checks during memory operations impose no additional costs. The SFITag analog also uses regular memory stores to simulate costs of \texttt{stg} but additionally augments all memory operations with an extra dummy load to simulate the additional memory bandwidth imposed by fetching tags.

We implement the HAKC analog by modifying \texttt{glibc}'s allocation function to use the analog in place of \texttt{stg} instructions. We implement the SFITag analog by additionally building an assembly rewriter tool (leveraging tooling infrastructure from LFI~\cite{lfi}) that inserts dummy loads after every memory operation in a given binary.

\para{Evaluation}
We test these analogs using the SPEC CPU 2006 benchmarks from \S\ref{subsec:spec} and compare the overheads from the analogs to actual hardware. The results (and associated error bars) are shown in Appendix Figures~\ref{fig:spec-analogs-pixel8}\ifextended{ and \ref{fig:spec-analogs-pixel9}}.

We find that neither analog accurately tracks real overheads and cannot usefully predict which benchmarks are likely to have higher overheads across microarchitectures. The HAKC analog seems to predict near-zero overheads for all workloads, while the SFITag analog predicts arbitrarily high overheads even when overheads are negligible (e.g., SFITag predicts over a $3.5\times$ higher runtime for \texttt{456\_hmmer} on Pixel 8's Big core with MTE ASYNC, while the reality is that the overhead on real hardware is negligible).

\subsection{Leveraging MTE for fast isolation}
\label{subsec:cage}

Prior work~\cite{cage} has also investigated the performance of MTE SYNC to speed up WebAssembly sandboxing by measuring slowdowns of PolyBench/C on the three Google Pixel 8 Cores—the same as our measurement setup in \S\ref{sec:colorguard}. However, due to a bug in their implementation, they incorrectly report extremely good performance with MTE-based isolation in Wasm, reporting speedups of $1.037\times$, $1.051\times$, and $1.339\times$ on Wasm---rather than the large spikes in performance of up to \polybenchxmax that we see in our measurements.

To understand the discrepancy, we investigated their open artifact~\cite{cage-artifact} and discovered two implementation bugs.

\para{Bug 1} The MTE-aware Wasm compiler they implemented (a fork of Wasmtime~\cite{wasmtime}) did not apply the \texttt{PROT\_MTE} flag to any pages requested after the start of the application---rather, the flag was only applied to memory allocated at startup.

\para{Bug 2} Several of the initial memory pages of the application contained the applications static data (corresponding to an ELF file's \texttt{.data} section); Wasmtime loads this into memory as copy-on-write file-backed memory mappings for efficiency~\cite{wasmtime-cow}. Unfortunately, such file-backed mappings do not support the \texttt{PROT\_MTE} flag\footnote{Their implementation specifies this flag, but the system call fails as tags on file-backed memory is not supported.}.

As a result of these two bugs, very few memory pages of the sandboxed application actually enforced MTE checks. 
For instance, their toolchain only tagged 16 of the total 784 memory pages in PolyBenchC's cholesky benchmark correctly with the \texttt{PROT\_MTE} flag.
This would account for the much better performance reported in their paper.

\para{Additional sources of overhead}
Even if the above bugs are fixed, this system~\cite{cage} as well as some of the other systems that leverage MTE for sandboxing~\cite{chen2024limitations} are susceptible to under-reporting overheads. To see why, suppose one of these schemes is being relied on to sandbox some part of an application (e.g., to isolate a library). Because these systems exclusively rely on MTE tags to enforce isolation, all pages in a process (and not just in the sandbox) must be mapped with \texttt{PROT\_MTE} and incur the cost of MTE. Thus, in our example, the cost of sandboxing is not just the MTE overheads imposed on the sandboxed library, but on the entire application as well.

Importantly, this distinction does not apply to the ColorGuard~\cite{segue-cg, segue-plas} use case evaluated in \S\ref{sec:colorguard}, which assumes only sandbox memory has the \texttt{PROT\_MTE} flag and incurs overheads. The reason being, with ColorGuard, the finer-grained MTE isolation is layered on top of coarse-grained SFI isolation, which isolates the rest of the process from the sandboxes.

\subsection{Improving performance by reusing tags}
\label{sec:tag-checking-slow}

Despite some gains from selective tagging for large-allocations (\S\ref{sec:selective-tagging}), substantial overhead persists on Pixel 8 Big and Perf core.
Prior work from Gorter et.al~\cite{gorter2024sticky} indicated that one way to recover performance was to reuse tags across re-allocations. Since they tested this approach on MTE ASYNC on the performance core of the Pixel, we check if their approach avoids the overheads of MTE ASYNC we saw on the Big core (\S\ref{sec:async-big-core-analysis}).
We expect that if our conclusions in \S\ref{sec:async-big-core-analysis} are correct, this would not change overheads much.

We repeated the \texttt{gcc} benchmark from \Cref{fig:glibc}, but used a modified allocator that tags all memory pages used by the benchmark only once during program initialization. We ran this benchmark on the Pixel's Big core and found that this was able to account for only $1.14\times$ of the \glibcasyncgccbig overheads we see on this benchmark.

\para{Interpretation}
Reusing tags offers a concrete performance improvement, consistent with the claims of prior work~\cite{gorter2024sticky}.
However, in the context of the large overheads we see in ASYNC performance on Pixel's Big core, we conclude that re-tagging is not the dominant source of overhead in these extreme cases such as \texttt{403.gcc}.
\section{An MTE Wishlist}
\label{sec:wishlist}

\para{More precise MTE ASYNC reporting} As discussed in \S\ref{sec:mte}, ASYNC MTE currently sacrifices precise enforcement and reporting in exchange for relaxed handling of the data dependency between a tag-load/tag-check and a load/store instruction. Given Ampere's results with SYNC, it's not clear that this relaxation is even required for performance. However, in the cases where implementers do opt to support ASYNC, providing precise reporting (i.e., saving the PC of the faulting instruction in a register) would greatly increase the odds that detected bugs will be rapidly diagnosed and fixed, while still providing greater freedom for implementers than SYNC.

Another relatively cheap way to simplify diagnosing the cause of ASYNC traps would be delivering an interrupt (imprecise trap) when a tag mismatch occurs. Even though these could potentially be thousands of cycles later given the size of modern instruction windows, the trap handler would still have a much fresher view of the system state than if the fault is not reported until the next system call, potentially making it easier to identify the faulty loop, function, etc., where the bug is, and offering the possibility of a fresher core dump.

\para{Logging tag mismatches} Our fast memory tracer could be faster and more versatile if tag mismatches were cheaper to record. At present, MTE SYNC traps on every mismatch. A trap per-event is very expensive---including running an interrupt handler and a full pipeline flush, which is increasingly costly on modern out-of-order processors. A new MTE mode where the precise PC and operands of the current instruction could be written to a buffer on a tag mismatch (similar to Intel PT~\cite{intel-manual}) would significantly accelerate tracing. Per our earlier point, since this only requires precise reporting, it could also be supported with MTE ASYNC.

\para{Revocation in multi-threaded programs}
While BufLock offers a sound solution for TOCTOU in re-entrant environments, it would not be safe in multi-threaded programs unless we also incorporate expensive \texttt{DMB} fences after tagging. Adding such synchronization barriers obviates any performance benefits vs. copying. This is unfortunate, because revocation is a general idea that is useful in a variety of contexts beyond just BufLock, including in the enforcement of memory safety~\cite{xia2019cherivoke}. Including dedicated instructions for synchronized tag updates in the MTE standard may allow hardware vendors to provide more optimized instructions for this purpose.

\section{Conclusion}

A deeper understanding of MTE's performance in real hardware is necessary to provide a solid foundation for both future hardware implementations and the software that uses it. As we have seen, there are a variety of implementation pitfalls that can hamstring MTE's performance and limit its deployment. However, this should not be viewed as a recommendation against MTE; rather our analysis offers guidance on avoiding such pitfalls. MTE also offers a variety of new opportunities for performance optimization in software systems. We look forward to the next generation of memory tagging systems that incorporate the lessons of prior generations and hope our analysis contributes to this effort.

\section*{Acknowledgments}
We thank the reviewers for their insightful feedback, Carl Worth for help developing the Linux kernel patch, Kostya Serebryany for feedback on early drafts of this work, Ampere for providing access to a bare-metal cloud server, NSF grants \#2327337 and \#2212579, and Google, Qualcomm, Mozilla for research gifts that helped support this work.

\section*{Ethical Considerations}
Our work focuses on understanding and improving the performance of hardware support for
a security feature, namely ARM MTE. We performed all experiments on local systems and did not use any systems with personal data. Concretely, we evaluate ARM MTE performance on the Google Pixel 8 and 9, and Ampere's \textit{AmpereOne} chip. They are commercially available~\cite{GoogleP8Announcment,ampereone-oracle} devices open for anyone to use.

While ARM MTE is focused on enforcing security, our analysis was limited to analyzing its performance, and did not focus on finding (or find) any security-related bugs that required reporting.
Our work didn't involve human subjects.

We reported our performance findings to Ampere and Google, both of whom acknowledged our report, viewed our findings positively, and passed on the report to internal teams. In Ampere's case, we also received a note about this being fixed in the next generation of CPUs which we have included in the paper. Both Google and Ampere are aware that all of our findings will be published at academic conferences and stated they don't have any concerns about this. The following lists entities that are possible stakeholders.
\begin{CompactItemize}
    \item This paper's authors and author affiliated institutions.
    \item All ARM CPU vendors, specifically Google and Ampere.
    \item Any customers of ARM chips interested in MTE, especially from Google or Ampere.
    \item Maintainers of MTE software support in the Linux kernel and libc.
    \item Developers of memory tagging standards on other platforms like RISC-V~\cite{riscvmte} and x86~\cite{x86chktag}. 
    \item Authors of the most related prior work, whose work we have cited, reproduced, or extended. In particular, authors of the papers.
    \begin{enumerate}
        \item "Sticky tags: Efficient and deterministic spatial memory error mitigation using persistent memory tags"~\cite{gorter2024sticky}
        \item "Cage: Hardware-accelerated safe WebAssembly"~\cite{cage}
        \item "Memory tagging and how it improves C/C++ memory safety"~\cite{kostya-tagging}
        \item "Segue\&Colorguard: Optimizing SFI performance and scalability on modern architectures"~\cite{segue-cg}
        \item "Preventing kernel hacks with HAKC"~\cite{hakc}
        \item "Sfitag: Efficient software fault isolation with memory tagging for ARM kernel extensions"~\cite{sfitag}
    \end{enumerate}
\end{CompactItemize}

\para{Potential negatives}
Since we show that MTE has overheads in some implementations and workloads, our results could be (mis)-interpreted in several scenarios

\begin{CompactItemize}
\item Hardware vendors may decide not to deploy this security extension in products in the near term. This would mean that an important security extension is not available to customers, reducing the effective security of consumer devices. 
\item Software vendors may delay enabling MTE on their products, which can result in software being insecure and attackable for longer. This may be exacerbated if they incorrectly assume that the worse performance overheads apply to them without actually measuring their software.
\item Software vendors may also inadvertently assume the opposite---that they have low MTE overheads as shown in some of our workloads and choose to deploy MTE without careful analysis. This would be an issue if their application falls into a case that has high overheads as shown in the paper or is a usecase we have not measured.
\item Users may chose to disable MTE on their devices at the system level as they worry about overheads.

\end{CompactItemize}

The net effect would nevertheless mean that an important security extension is not available to customers, reducing the effective security of consumer devices.

\para{Mitigations}
While the above scenarios are possible, we have taken great care to specify which workloads are affected as well as the context in which these results apply to minimize probability of misinterpretation. We have also noted this directly in our conclusion.
Our research, especially our observations on how to fix performance cliffs, methodology on how to evaluate performance for the next generation of MTE chips, provides a valuable resource as it outlines a path to address performance issues. The positive outcome of this research would be hardware and software vendors working together to increase adoptability of this important security technology. 

We believe our work clarifies the state of MTE, as well as the path forward to improve performance (and thus adoption and security of end user devices), by offering a clearer understanding of the costs and benefits.
On balance, we believe the net benefit to hardware vendors and software applications outweigh the potential but low-probability risks of misinterpretation or slowing down adoption in the short term. 

\section*{Open Science}  
\label{sec:open-science}

As part of our commitment to transparency and reproducibility of our work, we have released all tools developed, source code for the tools, build scripts, benchmark scripts and results under an open-source license. All artifacts are available at \anon{\url{https://github.com/anonymous-user1112/anon-mte-artifact}}{{\url{https://github.com/UT-Security/mte-root} and \formatdoi{10.5281/zenodo.17953065}}}.

Our artifact is divided into the following directories:
\begin{CompactItemize}
    \item \path{mte-playground}: Contains the scripts, data, and config files required for the SPEC benchmarks (\Cref{sec:perf,,sec:prior-misconceptions}; \Cref{fig:glibc,,fig:ampere-glibc-spec06-stlfwd-disabled,,fig:spec-analogs-pixel8\ifextended{,,fig:spec-analogs-pixel9,,fig:spec-analogs-ampereone}}).
    \item \path{mte-root}: Contains data from BufLock, MemTrace, server benchmarks, and all microbenchmarks discussed in the paper.
    \item \path{mte-server}: Includes the custom Phoronix test suite and profiles used for MTE server benchmarks (\Cref{subsec:server}; \Cref{fig:ampere-server-fix,,fig:ampere-memcached\ifextended{,,fig:ampere-server-bench-other}}).
    \item \path{mte-bench-clean}, \path{mte-performance}, \path{mte_benchmarks}: These directories contain the MTE ASYNC microbenchmark (\Cref{sec:async-big-core-analysis}; \Cref{fig:glibc-big-core}) and additional evaluative microbenchmarks \ifmain{(\Cref{appendix:extra-microbench}).}
    \ifextended{(\Cref{appendix:extra-microbench}; \Cref{fig:inst-counts,,fig:stg,,fig:micro-result}), respectively.}
    \item \path{mte-polybench}: Contains the ColorGuard-MTE implementation alongside the Polybench/C benchmark (\Cref{sec:colorguard}\ifextended{, \Cref{fig:polybench-all}}).
    \item \path{firefox-mte}: Contains BufLock Implementation, the TOCTOU protections for Firefox (\Cref{sec:toctou}; \Cref{tbl:buflock}).
    \item \path{mtetrap}: Contains MemTrace Implementation, the fast data tracing tool with MTE (\Cref{sec:trace}; \Cref{fig:memtrace-p8x\ifextended{,,fig:memtrace-all}}).
    \item \path{llvm-mte}, \path{mte-glibc}: These contain the TagCFI implementation and the custom glibc. The glibc is used for analogs as well (\Cref{sec:cfi,sec:prior-misconceptions,,appendix:cfi}\ifextended{; \Cref{fig:fwd-cfis-clang,,fig:bwd-tagcfi,fig:ampere-bwd-cfi}}).
\end{CompactItemize}
%


{
\bibliographystyle{abbrv}
\bibliography{paper}
}

\appendix
\section*{Appendix}

\section{Detailed algorithm for identifying structural hazards}

In \S\ref{sec:async-big-core-analysis}, we analyzed the root cause of the performance drop of MTE ASYNC in the Big core. The algorithm for the microbenchmark used is presented in Algorithm~\ref{alg:ll-bench}.

\begin{algorithm}
\SetKwInOut{Input}{input}\SetKwInOut{Output}{output}
\Input{$\textsf{Array-Length}$, $\textsf{LL-Length}$,  $\textsf{Stride}$}
\DontPrintSemicolon
\caption{Algorithm for microbenchmark to detect hazards used in \S\ref{sec:async-big-core-analysis}. }
\label{alg:ll-bench}
{$L$ = \textsf{LL-Length}}\;
{$A$ = \textsf{Array-Length}}\;
{$S$ = \textsf{Stride}}\;
{Construct a $\linklist$ $ll$ of size $L$}\;
\For{Element $el \in ll$}{
    {$el$.array = $\malloc(A)$}\;
    {fill-with-random-values($el$.array)}\;
}
{Evict $ll$ from cache}\;
{\textsf{Head} =  $ll$.head}\;
{Time the while loop below after warming it up}\;
\While{\textsf{Head} $\neq \perp$}{
    {\textsf{Sum} = 0}\;
    \For{$i = 1$, $i<A$, $i{+}{S}$}{
        {\textsf{Sum} = \textsf{Sum} + \textsf{Head}.array[i]}\;
    }
    {\textsf{Head} = \textsf{Head}.next}\;
}
\end{algorithm}

\section{Server benchmark setup}
\label{appendix:server}

In \S\ref{subsec:server}, we evaluate performance of server workloads: Nginx, RocksDB, PostgreSQL, and Memcached. We evaluate performance when MTE is enabled, normalized against when MTE is disabled. These benchmarks are repeated 4 times, and is allowed to use a different number of cores each time: 8, 32, 64, and 96 cores. Following Phoronix, each of benchmark is iterated a minimum of 3 times, and is repeated until stable up to a maximum of 15 times.

\para{Nginx}
Performance of the Nginx web server serving static files is measured for different number of incoming connections, where Nginx is allowed to automatically choose the number of worker threads.

\para{RocksDB}
Performance of the RocksDB database is measured by the throughput seven different database access patterns (e.g., random read, sequential fill etc.) 

\para{PostgreSQL}
Performance of the PostgreSQL database is measured with \texttt{pgbench}, which can test the database for different sizes of the database, different number of clients, worker threads, for both read-only and read-write SQL workloads.
Additionally, the maximum number of connections is limited to 6000. Shared memory buffer size is set to 8GB.

\para{Memcached}
Performance of the Memcached in-memory key-value store is measured by evaluating performance on a workload generated by Memtier~\cite{memtier}. Performance is measured for different values of Memcached threads, Memtier threads (where each thread has 50 clients making requests), for different ratios of set and get operations. Additionally, the maximum number of connections is limited to 4096.

\section{Additional microbenchmarks}
\label{appendix:extra-microbench}

\ifmain{This section is available in our extended paper~\cite{mte-extended}}
\ifextended{
We executed a number of microbenchmarks to understand the performance of MTE hardware. While these were not used to directly explain the overheads we saw in different benchmarks, they were informative to our understanding of the implementations. 

\subsection{Instruction cycles}

We measured the performance of individual instructions in Figure~\ref{fig:inst-counts} to help us identify the sources of different overheads. 

\begin{figure}
\centering
\begin{threeparttable}
\footnotesize
\setlength{\tabcolsep}{2pt}
\begin{tabular}{l ccccccc}
\hline
\textbf{Instruction} & \textbf{X3} & \textbf{A715} & \textbf{A510} & \textbf{X4} & \textbf{A720} & \textbf{A520} & \textbf{AmpereOne} \\
\hline
\textbf{\makecell{Clock speed\\(GHz)}} & 2.35 & 1.94 & 1.41 & 2.49 & 2.12 & 1.68& 2.59 \\
\hline  
\multicolumn{7}{l}{\textbf{Loads}} \\  
ldr    & 2.98 & 2.64  & 1.01 & 2.99 & 2.38 & 2.00  & 2.00 \\
ldr\tnote{*} & 2.98 & 2.64  & 1.03 & 2.99 & 2.38 & 1.99  & 2.00 \\
ldr +async  & 3.00 & 2.63  & 0.92 & 2.99 & 2.38 & 1.99  & 2.00 \\
ldr\tnote{*} +async & 3.00 & 2.64 & 1.02 & 2.99 & 2.38 & 2.00 & 2.00 \\
ldr +sync   & 2.98 & 2.63  & 0.88  & 2.99 & 2.38 & 2.00 & 2.00 \\
ldr\tnote{*} +sync & 2.98 & 2.63  & 0.94 & 2.99 & 2.39 & 1.99 & 2.00 \\
\hline  
\multicolumn{7}{l}{\textbf{Stores}} \\  
str & 1.98 & 1.83  & 1.00  & 1.99 & 1.74 & 1.00 & 2.00 \\
str\tnote{*} & 1.98 & 1.85  & 1.00  & 1.98 & 1.74 & 1.00 & 2.00 \\
str +async  & 1.99 & 1.81  & 1.00  & 1.99 & 1.73 & 1.00 & 2.00 \\
str\tnote{*} +async & 1.99 & 1.81  & 1.00 & 1.99 & 1.72 & 1.00& 2.00  \\
str +sync   & 0.14 & 1.57  & 0.45   & 0.14 & 1.60 & 0.65& 2.00  \\
str\tnote{*} +sync & 1.97 & 1.84  & 1.00  & 1.98 & 1.75 & 1.00& 2.00  \\
\hline  
\multicolumn{7}{l}{\textbf{Tag Ops}} \\  
ldg                     & 2.91 & 1.90  & 0.94  & 3.00 & 1.93 & 0.84 & 1.95 \\
stg                     & 1.00 & 1.82  & 1.00  & 1.00 & 1.76 & 1.00 & 1.00 \\
st2g                    & 1.00 & 1.83  & 0.46  & 1.00 & 1.77 & 0.50 & 0.50 \\
stzg                    & 1.00 & 1.84  & 0.98  & 1.00 & 1.76 & 1.00 & 1.00 \\
stz2g                   & 0.33 & 1.84  & 0.45  & 0.33 & 1.75 & 0.50 & 0.50 \\
stgp                    & 1.00 & 1.68  & 0.98  & 1.00 & 1.55 & 1.00 & 1.00 \\
dcgva                   & 0.13 & 0.14  & 0.19  & 0.14 & 0.14 & 0.03 & 1.00 \\
\hline
\end{tabular}
\begin{tablenotes}
    \small
    \item[*] Instruction accesses a memory page allocated without \texttt{PROT\_MTE}
\end{tablenotes}
\end{threeparttable}
\caption{Instruction throughput measurements across cores.}
\label{fig:inst-counts}
\end{figure}

\subsection{Overheads of tagging operations}
\label{sec:stg}

\begin{figure}[t]
    \centering
    \begin{subfigure}{\linewidth}
        \includegraphics[width=0.9\linewidth]{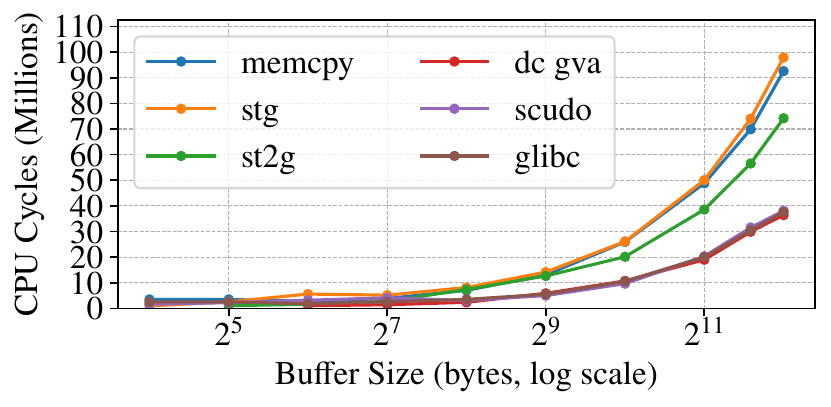}
        \caption{Pixel 8 Little Core}
        \label{fig:stg-p8l}
    \end{subfigure}
    \begin{subfigure}{\linewidth}
        \includegraphics[width=0.9\linewidth]{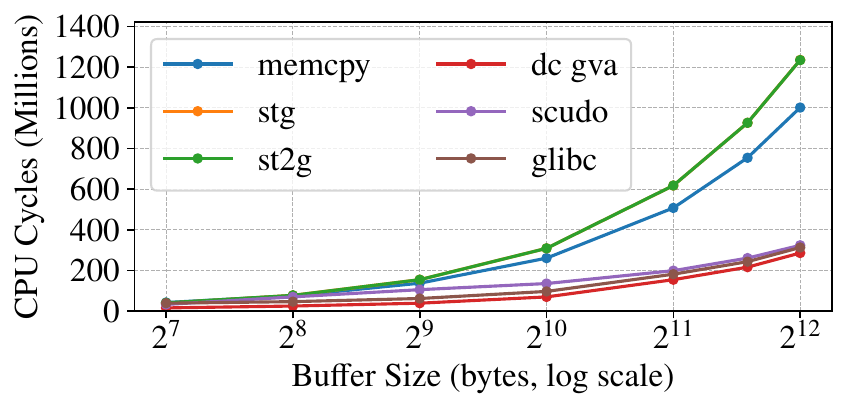}
        \caption{AmpereOne}
        \label{fig:stg-ampere}
    \end{subfigure}
    \caption{The average number of CPU cycles to tag a given buffer with different tagging primitives on the Little core of Pixel 8 (performance is very similar on other Pixel cores) and AmpereOne. We see that simply using tagging primitives in a loop is slow (as a reference, it is slower than a \texttt{memcpy}). However, glibc's and scudo algorithms that use a mix of these operations in unrolled loops significantly improve performance.}
    \label{fig:stg}
\end{figure}

For the use cases we consider such as BufLock~(\S\ref{sec:toctou}), we need to evaluate the performance of tagging primitives (like \texttt{stg} which tags 16 bytes of memory) vs. the performance of \texttt{memcpy} (since BufLock achieves its speedups by substituting tagging for data copies).
Fink et. al~\cite{cage} also ran a similar test to check the performance of different tagging primitives and compared this to the performance of \texttt{memset}, however, this doesn't quite help us understand BufLock.

\para{Evaluation}
We run a microbenchmark that allocates a buffer and measures the time taken to tag the buffer with each of these primitives 1000 times; we also flush the data cache on each iteration for consistent results.
We evaluate the MTE tagging primitives for different sizes: \texttt{stg} (16 bytes), \texttt{st2g} (32 bytes), \texttt{dcgva} (cache-line, i.e., 64-bytes)\footnote{MTE also provides \texttt{stgm} (variable size tagging), but this cannot be used by userspace applications, so we don't benchmark this}.
We compare this to the tagging approaches used in glibc and Android's Scudo libc; both libraries use a complex combination of \texttt{stg}, \texttt{st2g}, \texttt{dcgva} to efficiently tag buffers of different sizes.
Finally, we also compare the performance of raw \texttt{memcpy} of data, rather than tagging it.

\para{Interpretation}
We find that naively using one of the tagging primitives in a loop does not give optimal performance to tag a buffer of a given size; indeed it is slower than a simple \texttt{memcpy} of the same buffer. This implies that BufLock~\S\ref{sec:toctou} would be a pessimization if we tagged data with one of these primitives in a loop---something we tested and verified. Instead, for performant tagging, we should use the more careful combination of tagging instructions from glibc or Scudo, which are able to outperform \texttt{memcpy} on the buffer. Detailed data is available in the Appendix Figure~\ref{fig:stg}.

\subsection{Performance of different memory-access patterns}
\label{subsec:microbench}

We also constructed three microbenchmarks that simulate common memory access patterns: (1) read after read (RAR), (2) write after write (WAW), and (3) read after write (RAW).

\para{Setup}
All benchmarks were statically compiled using aarch64-linux-gnu-g++ 13.3.0 (GCC).
We enabled MTE on all mmap-allocated memory regions by specifying the \texttt{PROT\_MTE} flag. 
We report performance results as the relative slowdown of MTE SYNC and MTE ASYNC modes compared to the baseline execution without MTE tagging enabled.
Our hardware platform and setup selection is the same as that of~\Cref{e-setup}.

\para{Microbenchmark}
Our microbenchmark first uses \texttt{mmap} to allocate a 16 MiB \texttt{uint64\_t} main buffer, which has element count equals to $\texttt{len}=16*1024*1024/8$. 
The buffer is mapped under three configurations: without MTE, with MTE ASYNC, and with MTE SYNC. 
We initialize the buffer with a pointer-chasing pattern, where each \texttt{uint64\_t} element stores the address of another element within the buffer. 
This construction ensures that starting from the head element, it is possible to traverse the entire buffer by repeatedly following the stored addresses.
We then allocate an additional \texttt{uint64\_t} index buffer (without MTE) and initialize each element with a randomly selected location in the main buffer. 
By iterating over buffer[index[$i$]] with $i$ from $0$ to \texttt{len}, the benchmark accesses all elements of the main buffer in a randomized order.

\begin{itemize}
    \item RAR: In this microbenchmark, we begin from the head element of the main buffer and iteratively dereference each visited \texttt{uint64\_t} element. 
    The benchmark features the pointer-chasing memory access pattern and issues a sequence of dependent memory reads, where each access relies on the address obtained from the previous one. 
    \item WAW: In this microbenchmark, we iterate over buffer[index[$i$]] with $i$ from $0$ to \texttt{len} and perform write to each buffer[index[$i$]].
    The benchmark issues a sequence of independent memory writes, with each write targeting a random location in the buffer.
    \item RAW: In this microbenchmark, we iterate over buffer[index[$i$]] with $i$ from $0$ to \texttt{len} and, for each step, load a value from buffer[index[$i$]] and store it into buffer[index[$i+1$]].
    This access pattern implements store-to-load forwarding, where each memory write directly depends on the result of the prior memory read, and all memory accesses in buffer are to random locations.
\end{itemize}


\begin{figure}[]
\centering
\begin{tabular}{|l|l|l|l|}
\hline
                            & RAR         & WAW  & RAW  \\ \hline
Pixel 9 Little (ASYNC)      & $1.73\times$ & $1.00\times$    & $4.00\times$    \\ \hline
Pixel 9 Big (ASYNC)         & $1.17\times$ & $1.00\times$    & $1.00\times$    \\ \hline
Pixel 9 Performance (ASYNC) & $1.25\times$ & $1.20\times$  & $1.25\times$ \\ \hline

Pixel 8 Little (ASYNC)      & $1.16\times$ & $1.01\times$  & $4.90\times$    \\ \hline
Pixel 8 Big (ASYNC)         & $1.12\times$  & $1.16\times$ & $1.02\times$    \\ \hline
Pixel 8 Performance (ASYNC) & $1.19\times$ & $1.00\times$  & $1.00\times$  \\ \hline

Pixel 9 Little (SYNC)       & $1.76\times$ & $15.5\times$ & $14.0\times$   \\ \hline
Pixel 9 Big (SYNC)          & $1.16\times$ & $1.30\times$  & $1.50\times$  \\ \hline
Pixel 9 Performance (SYNC)  & $1.25\times$ & $2.80\times$  & $1.75\times$ \\ \hline

Pixel 8 Little (SYNC)       & $1.20\times$ & $4.61\times$  & $5.18\times$  \\ \hline
Pixel 8 Big (SYNC)          & $1.13\times$  & $1.67\times$    & $1.65\times$  \\ \hline
Pixel 8 Performance (SYNC)  & $1.19\times$ & $2.17\times$    & $1.15\times$ \\ \hline

Ampere (SYNC)               & $1.02\times$ & $1.80\times$  & $1.50\times$  \\ \hline
\end{tabular}
\caption{Relative performance overhead of Microbenchmarks}\label{fig:micro-result}
\end{figure}

\para{Results}
For each microbenchmark running each testing device, we compute and report the ratio of MTE SYNC over baseline and MTE-ASYNC over baseline in~\Cref{fig:micro-result}.
Across all devices, RAR incurs less than $2 \times$ overhead. 
However, WAW incurs significant overhead across almost all devices when running in SYNC mode, and RAW incurs significant overhead on Pixel Little core.

} 

\section{Using MTE to speed up CFI enforcement}
\label{appendix:cfi}

\ifmain{This section is available in our extended paper~\cite{mte-extended}.}
\ifextended{
In \S\ref{sec:cfi}, we evaluated the performance benefits of using MTE to speed up CFI. Our conclusion was that the performance of MTE-optimized CFI did not offer significant benefits over software CFI. We will provide more details on our MTE-enabled CFI schemes and their associated overheads.

\subsection{Forward-Edge CFI (VTT and IFCT)} \label{appendix:fwd-tagcfi}

Control-flow Integrity (CFI) restricts control flow to legitimate targets of indirect branches determined at compile time to mitigate code reuse attacks. As software CFI can be costly, processors have introduced branch target instructions for compilers to insert at valid indirect branch targets, such as x86-64's (endbr64)~\cite{intelcet} and ARM64's (\asm{bti})~\cite{armv9-manual}.

For forward-edge CFI, our approach called Tag-CFI, uses \kw{virtual table tagging} (VTT), to protect C++ virtual functions, and \kw{indirect function call tagging} (IFCT), to protect indirect function calls.

\begin{figure*}[t]
    \centering
    \begin{subfigure}[t]{0.5\linewidth}
         \centering
         \includegraphics[width=1.0\linewidth]{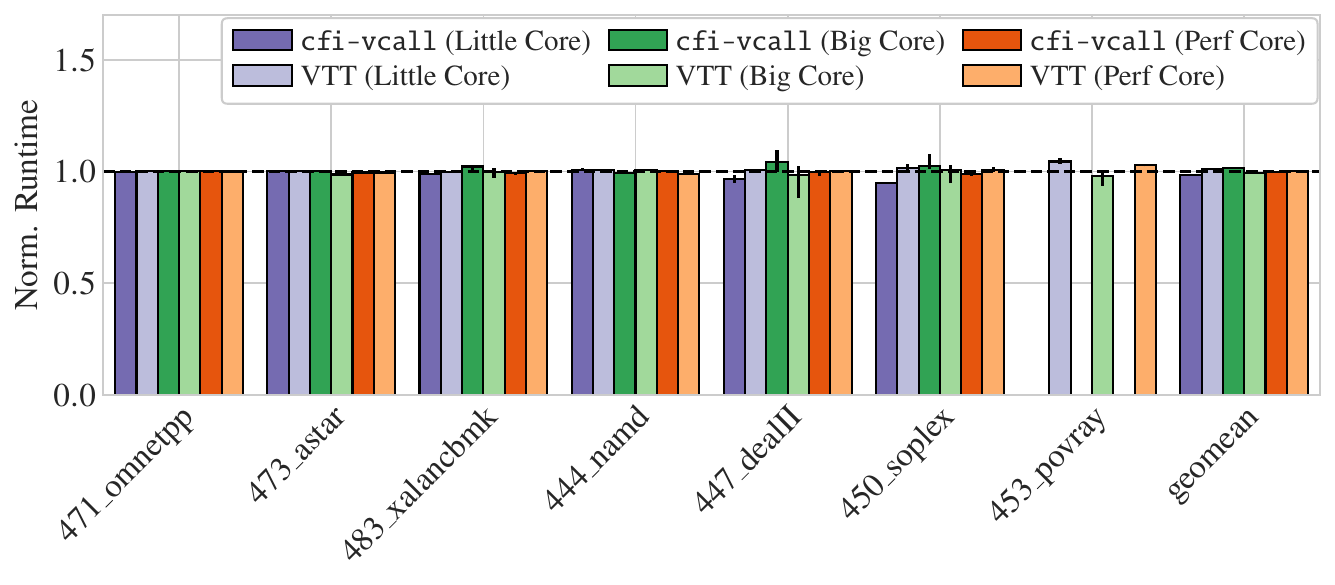}
         \caption{Protections for virtual table calls (VTT)}
         \label{fig:vtt-clang}
    \end{subfigure}%
    \begin{subfigure}[t]{0.5\linewidth}
         \centering
         \includegraphics[width=1.0\linewidth]{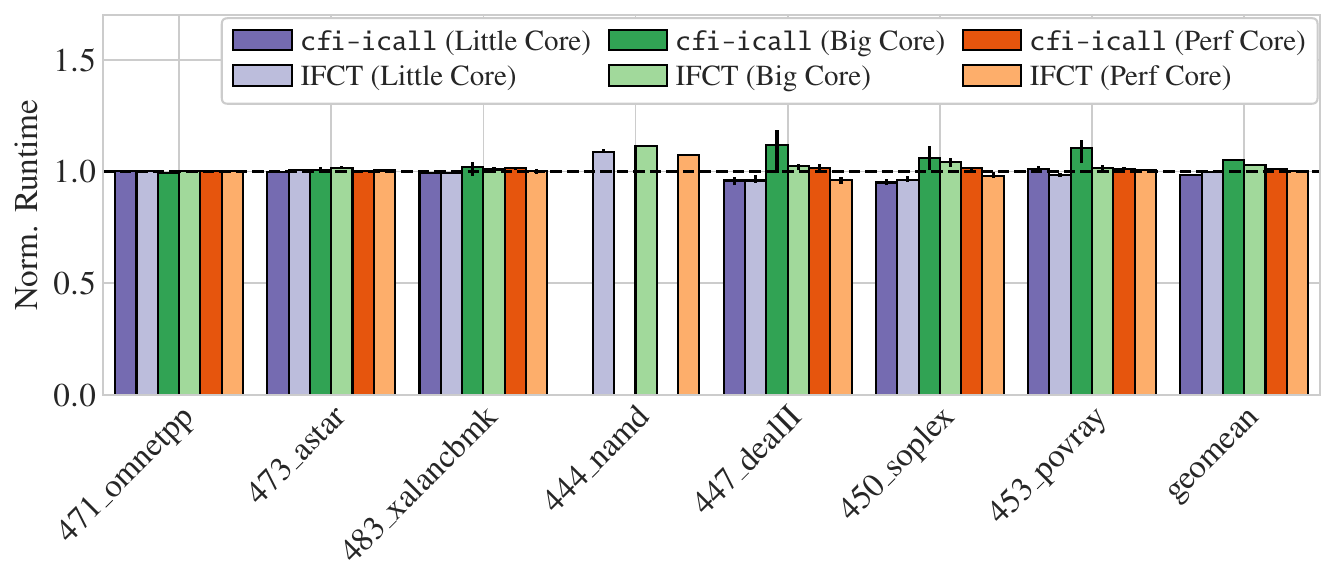}
         \caption{Protections for indirect function calls (IFCT)}
         \label{fig:ifct-clang}
    \end{subfigure}
    \caption{MTE-based and Clang's Software-based Forward-edge CFI schemes overheads: We compare the performance of TagCFI overheads on all C++ programs among SPEC CPU 2006 benchmark. TagCFI Performance numbers are normalized to the non-LTO baseline performance without any instrumentation and Clang CFI results are normalized to the LTO baseline performance without any instrumentation.}
\label{fig:fwd-cfis-clang}
\end{figure*}

\para{Virtual Table Tagging (VTT)}
Virtual tables (VTables) are a common target of memory safety exploits in C++ programs~\cite{gawlik2014towards}. Since attackers cannot modify VTables directly (they are stored in read-only memory), they instead exploit bugs to corrupt pointers to VTables in C++ objects.

To prevent these attacks, our scheme starts by: (1) reserving a tag exclusively for VTable pointers; (2) tagging all VTable memory with this tag; (3) masking the appropriate tag into legitimate pointers to VTables prior to access. Thus, pointers are restricted to accessing legitimate VTables.

To add finer-grained CFI to this, we: (1) store function equivalence class labels~\cite{abadi2009control}---a unique integer for each function signature---in the VTables next to the function pointers; (2) emit instrumentation before invoking virtual functions.

Finally, to prevent attacks that rely on misaligning the VTable pointer (e.g., incrementing it by 1) to read an incorrect FEC label as code, we force VTable pointers to be 16-byte aligned prior to label checks.

\para{Indirect Function Call Tagging (IFCT)} IFCT offers similar protections for indirect function calls (calls through function pointers) with the following steps: (1) IFCT modifies code generation to create a "jump table"---with the same function addresses and interleaved 4-byte labels format as VTT---populated with the address of any functions in the binary that have their address taken at any point in the code. (2) Any function pointer is replaced with an indexing address within this jump table, and any function pointer invocation uses this index to first retrieve the function address---with label and alignment checks similar to VTT---from the table.

\para{Evaluation}
We evaluated VTT and IFCT using C++ benchmarks in SPEC CPU 2006~\cite{spec2006}. We found that VTT shows that the Little core performance is slightly worse than the others, showing overhead of \vttlittlemax and \vttlittlegeomean for the worst performance and geomean, respectively. The results from the Big and Performance core stay within a geomean of \vttxgeomean as shown in Figure~\ref{fig:vtt-clang}. For IFCT in Figure~\ref{fig:ifct-clang}, its geomean overhead is under \ifctbiggeomean on all cores; the worst-case overheads are \ifctpovrayx–\ifctpovraybig on povray.
Additionally, we compare our scheme to Clang's CFI implementation where programs are compiled with \texttt{-fsanitnize=cfi-icall,cfi-vcall} options. Clang's CFI unfortunately introduces crashes in some programs in SPEC CPU 2006, so we exclude their performance from the geomeans for accurate comparison. We see that Tag-CFI is on par with or outperforms Clang's software-enforced CFI by a small margin, mostly on the Big core and the Performance core.

\subsection{Backward-Edge CFI (TagSS and RAL)}
\label{sec:TagSS}

Backward-edge CFI ensures that functions return to their original call site.
Historically, backward-edge protections were offered through a shadow stack---a
region of memory that stores copies of return addresses after call
instructions. These copies are either directly used as return addresses (as
they cannot be corrupted due to memory safety errors in the program) or checked
against the return address on the stack (to ensure this has not been corrupted).

Unfortunately, software implementations of shadow stacks must rely on
SFI/software sandboxing to isolate the shadow stack---an approach that can
impose a penalty around 7\% on the entire program~\cite{wahbe1993efficient, lfi}.
Thus, production implementations instead rely on
randomization to hide the shadow-stack---a best-effort approach that is not
secure.

ARM hardware also supports pointer authentication codes (PAC)---a mechanism which signs code pointers to protect them from tampering---an alternate way to protect return addresses by signing pointers.
Unfortunately, PAC does not prevent return address-reuse; an attacker can simply read a return address in some prior frame and overwrite the current return address on the stack with the prior address.

\begin{figure*}[t]
     \centering
\begin{subfigure}[t]{0.5\textwidth}
     \includegraphics[width=1.0\textwidth]{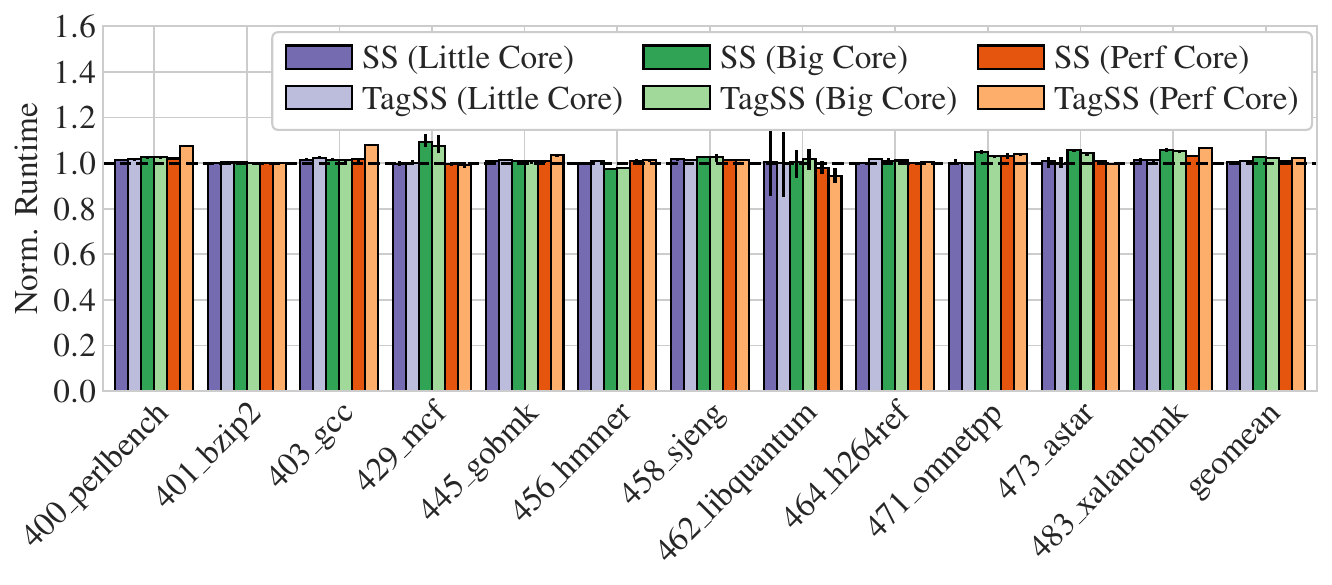}
     \caption{Protection for a shadow stack (TagSS)}
     \label{fig:tagss}
\end{subfigure}%
\begin{subfigure}[t]{0.5\textwidth}
     \includegraphics[width=1.0\textwidth]{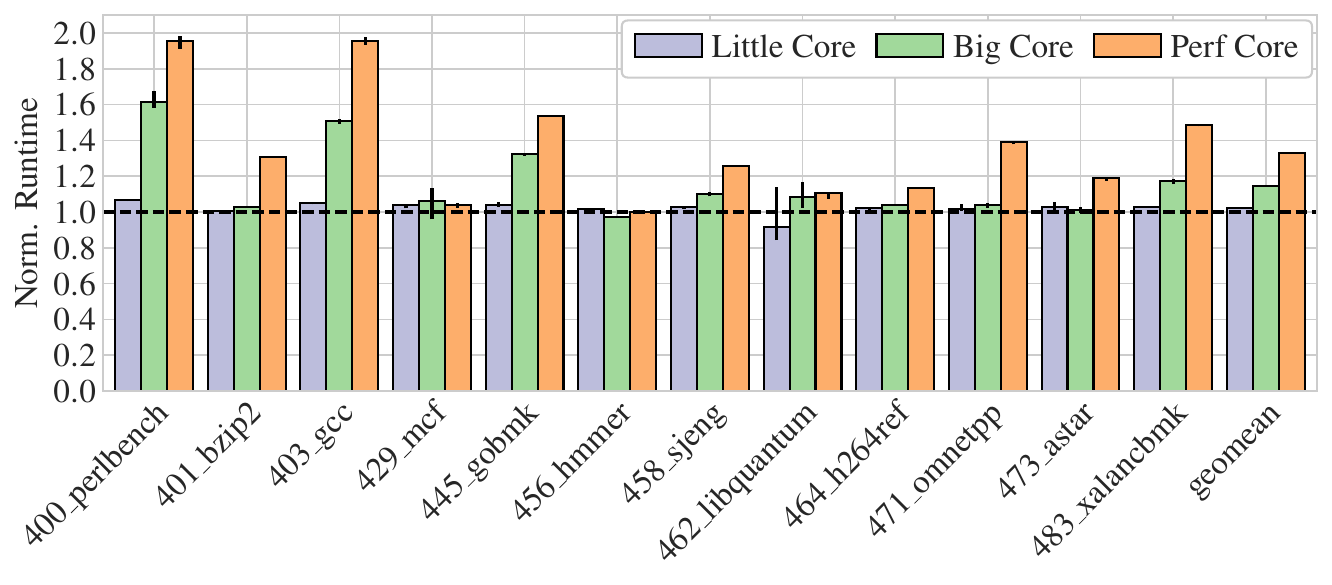}
     \caption{Protection for return address slots (RAL)}
     \label{fig:ral}
\end{subfigure}
\caption{Performance overhead of backward-edge TagCFI schemes. RAL Performance Overhead normalized to the baseline without any instrumentation. The Performance core shows high overhead due to serializing stores, and all workload with severe slowdown is stack-intensive workloads, such as \texttt{perlbench} and \texttt{gcc}. TagSS Performance Overhead compared to Clang ShadowCallStack (\texttt{-fsanitize=shadow-call-stack}). Its geomean is around 1\% and the worst performance stays within \tagsslittlemax across all cores.}
\label{fig:bwd-tagcfi}
\end{figure*}

\para{Tagged Shadow Stack (TagSS)}
TagSS tags the shadow stack with a dedicated MTE tag that is not used by any other part of the program; this ensures that the shadow stack can only be accessed by the software shadow stack instructions.
The program uses the shadow stack with a pinned register that holds a shadow stack pointer, \texttt{x18}.
We use Clang/LLVM's ShadowCallStack instrumentation pass~\cite{clangSS} that modifies function prologues to push the prior return address to the shadowstack along and pop the return address from the shadow stacks at the function epilogues.
Our TagSS runtime prepares a tagged shadow stack and set \texttt{x18} with a tagged pointer to the top of the shadow stack.

\para{Evaluation}
In Figure~\ref{fig:tagss}, we observe that TagSS imposes a geomean penalty of \tagsslittlegeomean, \tagssbiggeomean, \tagssxgeomean on the Little, Big, Performance core in order on the SPEC CPU benchmarks compared to Clang's ShadowCallStack~\cite{clangSS}.
TagSS achieves deterministically secure shadow stack with only small overheads.
Unfortunately, when comparing security with other hardware such as Intel CET~\cite{intelcet} or ARM GCS~\cite{armgcs}, TagSS does not offer a clear way to reserve a tag for its use. Hardware support allowing this could make this scheme more practical in the future.

\para{An alternate approach (RAL)}
Memory tagging offers another way to support backward edge CFI---return address locking (RAL). The intuition behind RAL is that a dedicated tag is used to tag the stack the frame-pointer and return-address stack-slots in each stack-frame. This ensures that the values can not be corrupted by other memory operations.

Figure~\ref{fig:ral} shows the performance overheads of RAL, and it imposes an overhead of \rallittlegeomean, \ralbiggeomean, \ralxgeomean for the Little, Big, and Performance core respectively. Although the performance core is slow (for reasons discussed in \S\ref{sec:serial_store}), RAL has reasonable overheads for the Little and the Big cores. Overall, RAL represents a trade-off: better code compatibility for more permissive backward-edge behavior.

Finally, we show the performance of backward-edge protections on AmpereOne in Figure~\ref{fig:ampere-bwd-cfi}, to demonstrate how overheads change on more efficient MTE implementations.

\begin{figure}[]
\centering
\includegraphics[width=1.0\linewidth]{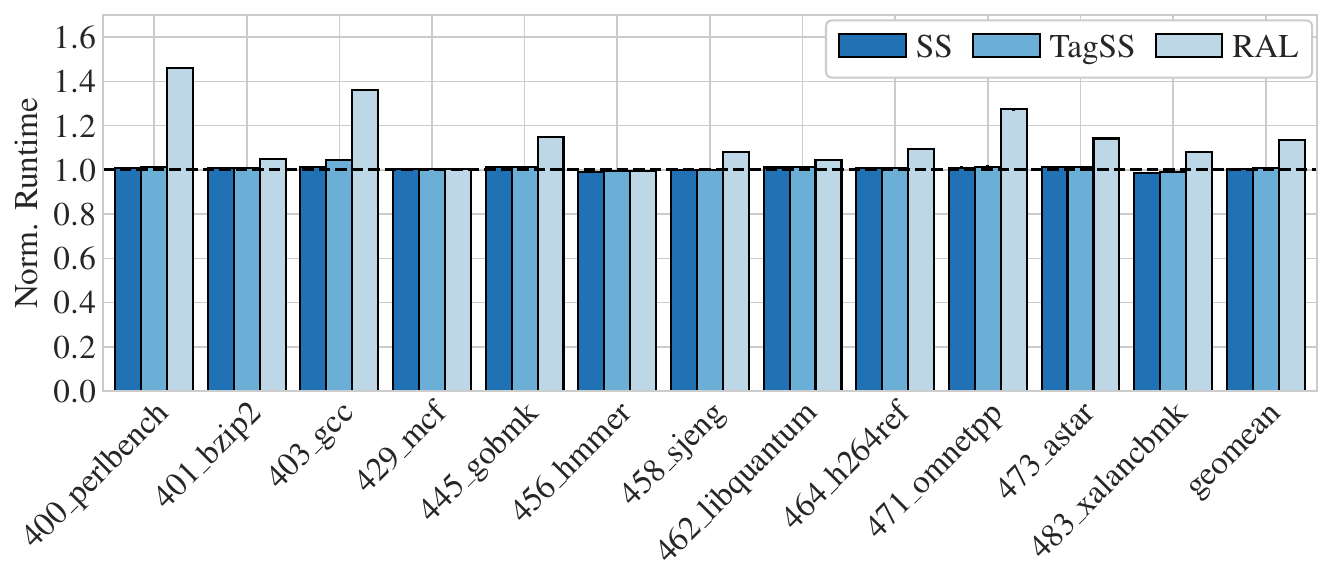}
\caption{Performance overhead of backward-edge TagCFI schemes tested on AmpereOne. RAL Performance Overhead normalized to the baseline without any instrumentation. TagSS Performance Overhead compared to Clang ShadowCallStack (\texttt{-fsanitize=shadow-call-stack}).}
\label{fig:ampere-bwd-cfi}
\end{figure}

} 

\section{Extra figures}
\label{sec:extra-figures}
This appendix section contains extra figures that complement our findings presented in the earlier sections. The following list maps each figure to its relevant section.

\begin{itemize}
    \item \textbf{Figure}~\ref{fig:ampere-glibc-spec06-stlfwd-disabled} shows SPEC 2006 overheads without the AmpereOne's store-to-load forwarding (which has MTE-related performance bugs) as discussed in \S\ref{sec:ampere-store-to-load}.

    \item \ifmain{\textbf{Figure}~\ref{fig:spec-analogs-pixel8} shows Pixel 8 results with software analogs, mentioned in \S\ref{subsec:analogs}, compared with hardware MTE results in \Cref{fig:glibc}.}
    \ifextended{\textbf{Figure}~\ref{fig:spec-analogs-pixel8}, \ref{fig:spec-analogs-pixel9} and \ref{fig:spec-analogs-ampereone}: These three figures include every data point in \Cref{fig:glibc}, and additionally have results from software analogues, mentioned in \S\ref{subsec:analogs}.}

    \ifmain{
    \item \textbf{Additional Server Benchmarks:} MTE overhead results for three additional server applications without the patch discussed in \S\ref{subsec:server} are provided in our extended paper~\cite{mte-extended}.

    \item \textbf{MemTrace Results on other architectures} Full MemTrace data (\S\ref{sec:trace}) for other cores---specifically the Pixel 8 (Little and Big) and AmpereOne---are provided in our extended paper~\cite{mte-extended}.

    \item \textbf{Polybench Evaluation:} The performance data of the Polybench/C suite compiled with the wasm2c-MTE---a WebAssembly compiler leveraging MTE (\S\ref{sec:colorguard})---is included in our extended paper~\cite{mte-extended}.
    }
    \ifextended{
    \item \textbf{Figure}~\ref{fig:ampere-server-bench-other} shows MTE overheads from other three server application before applying our patch, mentioned in \S\ref{subsec:server}.
    
    \item \textbf{Figure}~\ref{fig:memtrace-all} includes additional results of MemTrace on other cores; Pixel 8 Little, Big Core and AmpereOne mentioned in \S\ref{sec:trace}.
    
    \item \textbf{Figure}~\ref{fig:polybench-all} shows the performance polybench/C benchmark suite when compiled with wasm2c-MTE---a WebAssembly compiler leveraging MTE discussed in \S\ref{sec:colorguard}.
    }
    
\end{itemize}

\begin{figure}[h]
    \centering
    \begin{subfigure}{\linewidth}
        \centering
        \includegraphics[width=\linewidth]{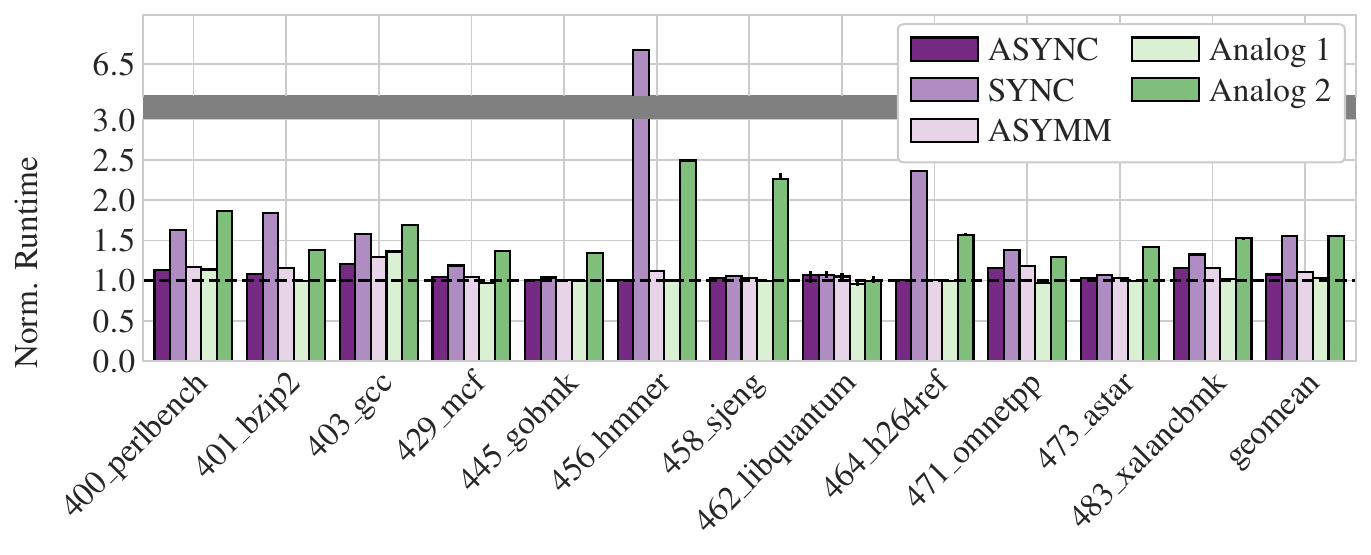}
        \caption{Pixel 8 Performance Core}
        \label{fig:pixel8-perf-glibc}
    \end{subfigure}
    \begin{subfigure}{\linewidth}
        \centering
        \includegraphics[width=\linewidth]{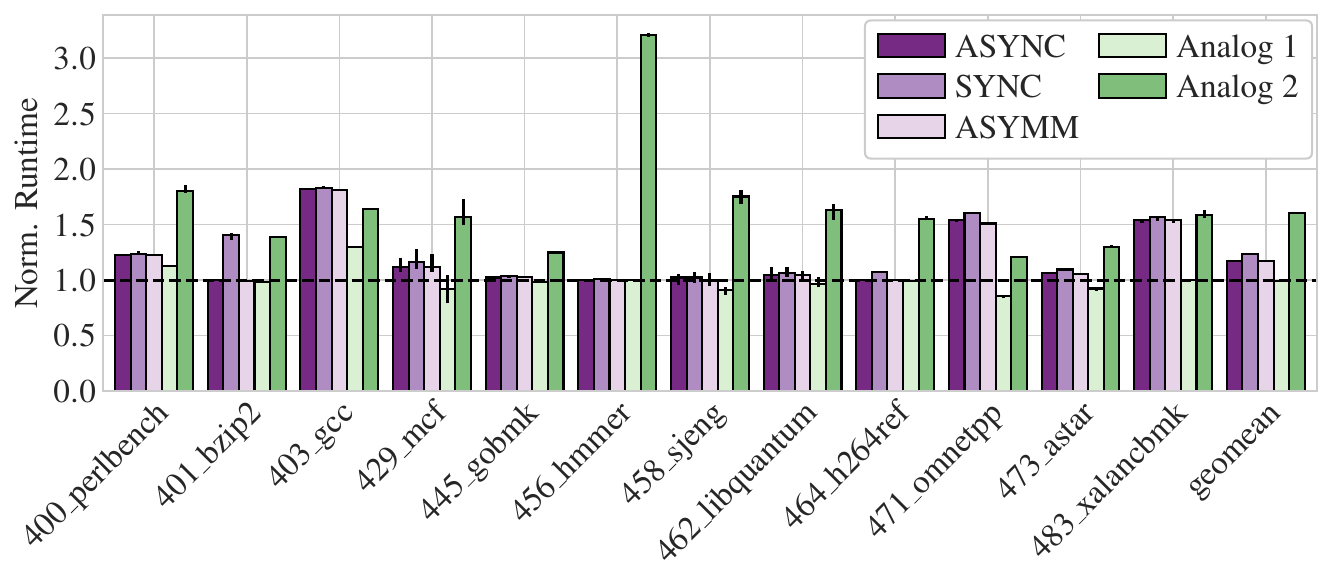}
        \caption{Pixel 8 Big Core}
        \label{fig:pixel8-big-glibc}
    \end{subfigure}
    \begin{subfigure}{\linewidth}
        \centering
        \includegraphics[width=\linewidth]{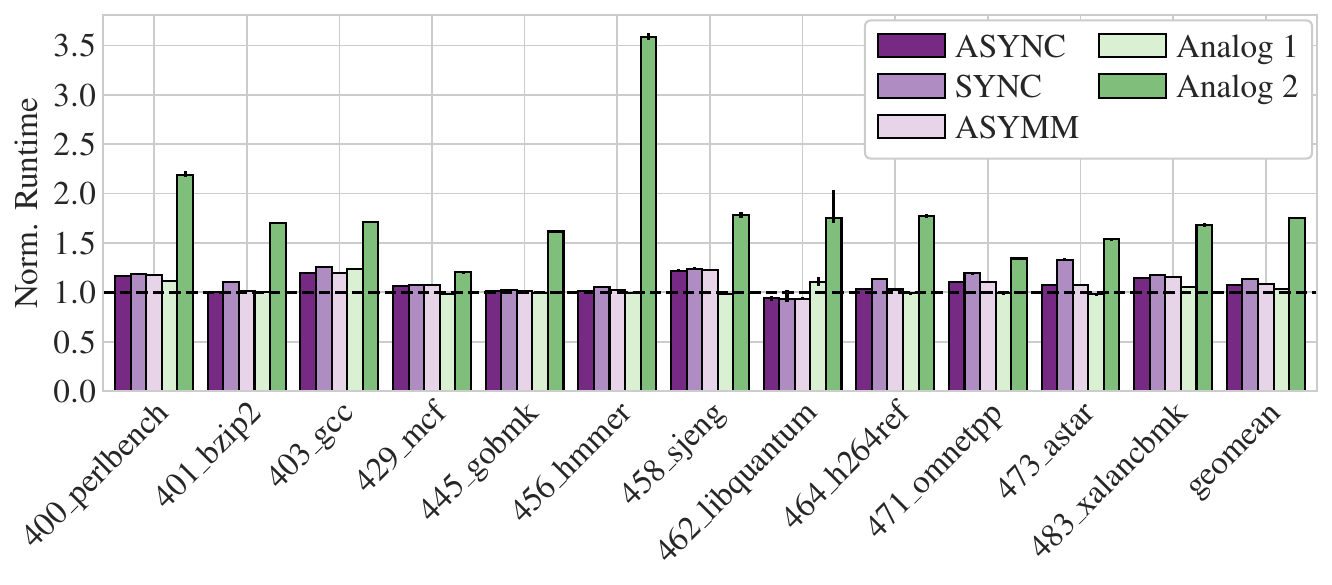}
        \caption{Pixel 8 Little Core}
        \label{fig:pixel8-little-glibc}
    \end{subfigure}
    \caption{Performance overhead across Pixel 8 cores for two prior MTE analogs, HAKC~\cite{hakc} (``Analog 1'') and SFI-Tag~\cite{sfitag} (``Analog 2''), compared with three hardware MTE modes (SYNC, ASYNC, ASYMM).}
    \label{fig:spec-analogs-pixel8}
\end{figure}

\begin{figure}[]
\centering
\includegraphics[width=\linewidth]{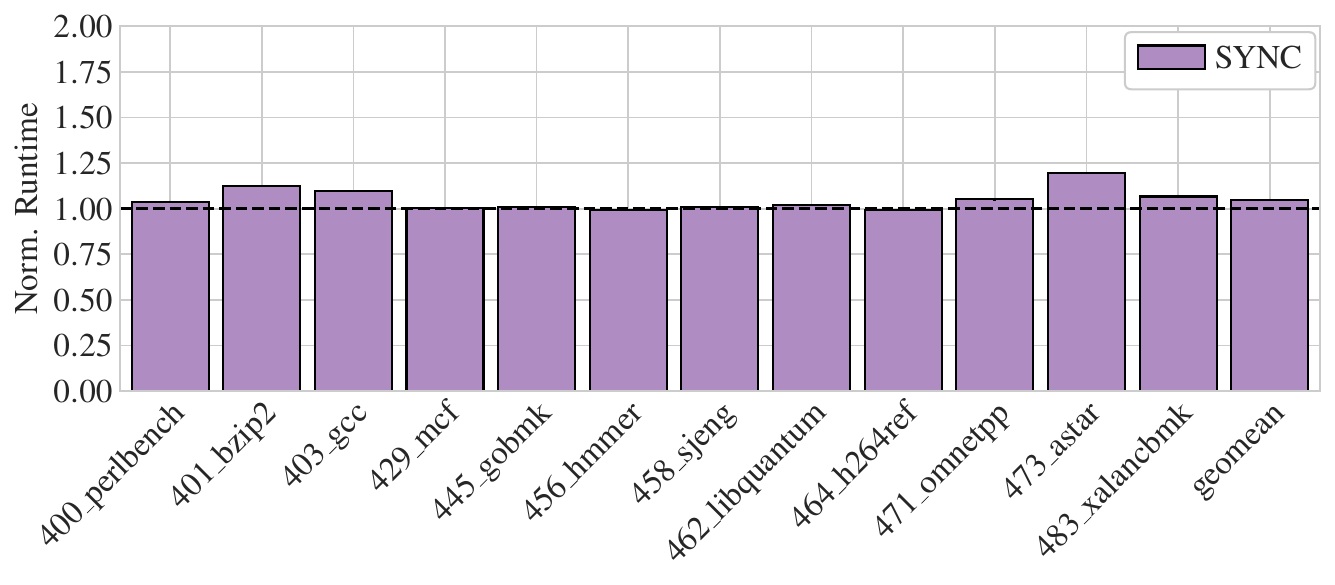}
\caption{Relative overhead of MTE-hardened allocator with \textbf{Store-to-Load forwarding disabled} on CPU2006 INT benchmark. Overhead on \texttt{456.hmmer} is reduced from \amperehmmer to negligible overhead, however a new bottleneck emerged on \texttt{473.astar} which increased from \ampereastar to \ampereastarstldisabled. Overall geomean of overhead dropped from \amperespecgeomean to \amperespecgeomeanstldisabled when removing store-to-load forwarding variability.}
\label{fig:ampere-glibc-spec06-stlfwd-disabled}
\end{figure}

\ifextended{
\begin{figure}[]
    \centering
    \begin{subfigure}{\linewidth}
        \centering
        \includegraphics[width=\linewidth]{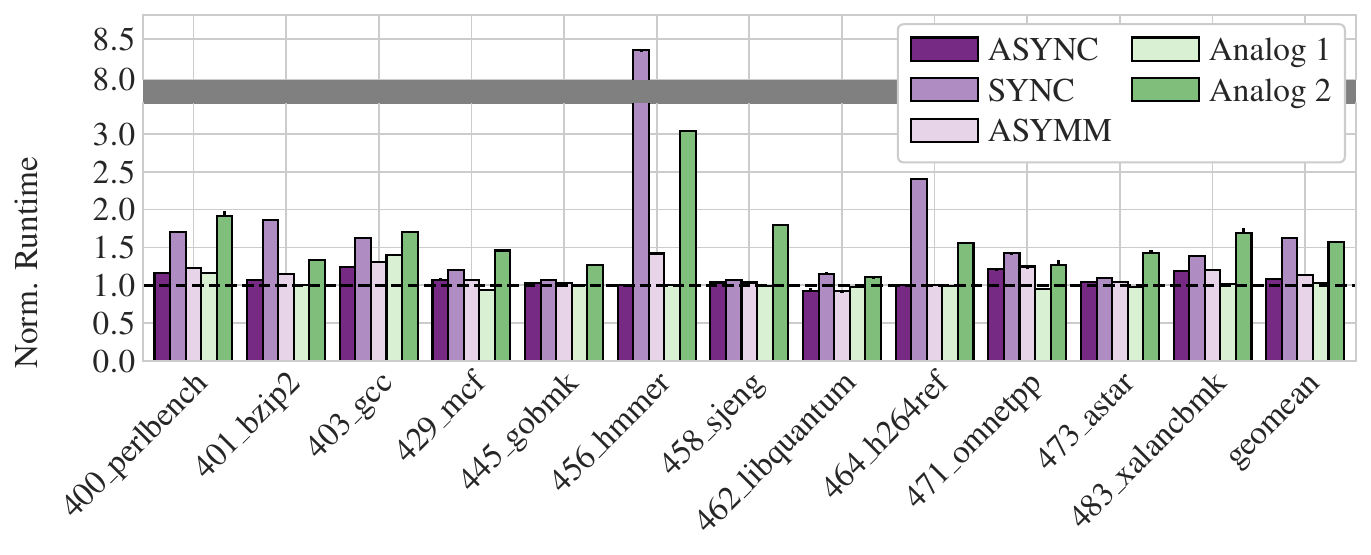}
        \caption{Pixel 9 Performance Core}
        \label{fig:pixel9-perf-glibc}
    \end{subfigure}
    \begin{subfigure}{\linewidth}
        \centering
        \includegraphics[width=\linewidth]{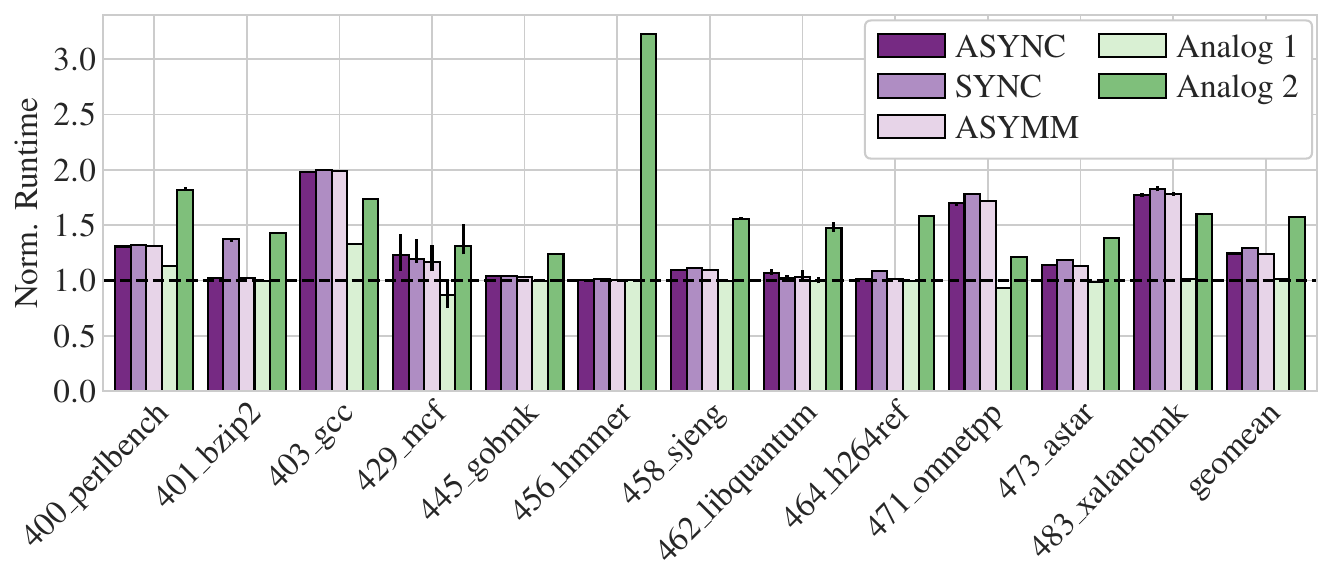}
        \caption{Pixel 9 Big Core}
        \label{fig:pixel9-big-glibc}
    \end{subfigure}
    \begin{subfigure}{\linewidth}
        \centering
        \includegraphics[width=\linewidth]{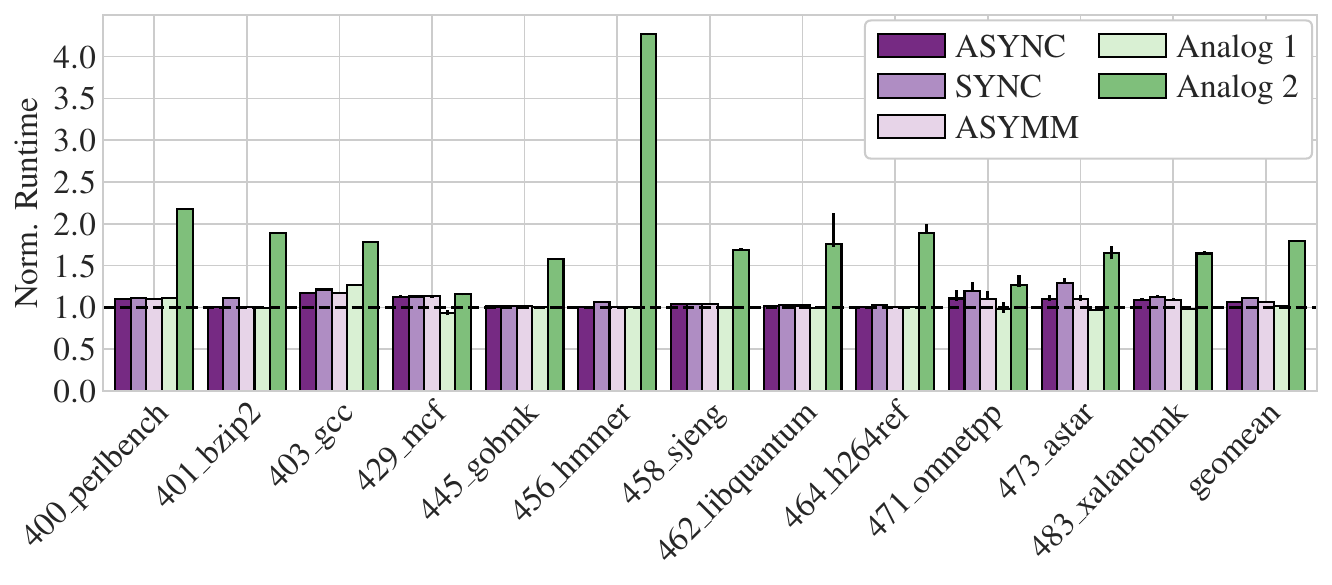}
        \caption{Pixel 9 Little Core}
        \label{fig:pixel9-little-glibc}
    \end{subfigure}
    \caption{Performance overhead across Pixel 9 cores for two prior MTE analogs, HAKC~\cite{hakc} (``Analog 1'') and SFI-Tag~\cite{sfitag} (``Analog 2''), compared with three hardware MTE modes (SYNC, ASYNC, ASYMM).}
    \label{fig:spec-analogs-pixel9}
\end{figure}

\begin{figure}[]
    \centering
    \includegraphics[width=\linewidth]{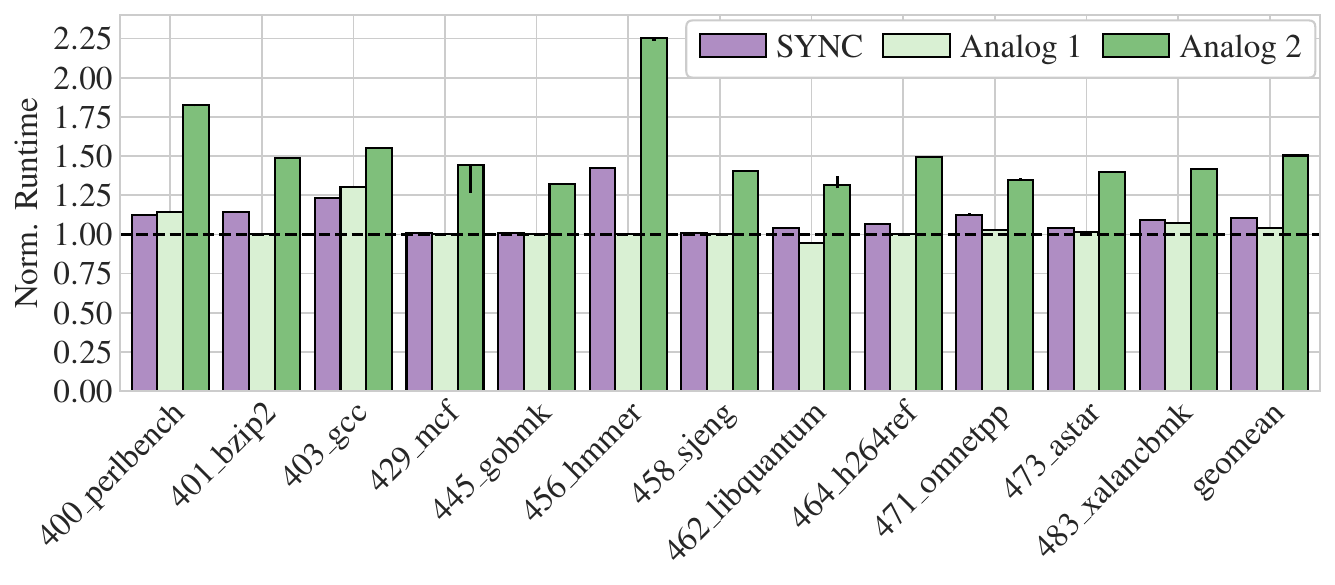}
    \caption{Performance overhead on AmpereOne CPU for two prior MTE analogs, HAKC~\cite{hakc} (``Analog 1'') and SFI-Tag~\cite{sfitag} (``Analog 2''), compared with its SYNC mode.}
    \label{fig:spec-analogs-ampereone}
\end{figure}

\begin{figure}
    \centering
    \begin{subfigure}{\linewidth}
        \centering
        \includegraphics[width=\linewidth]{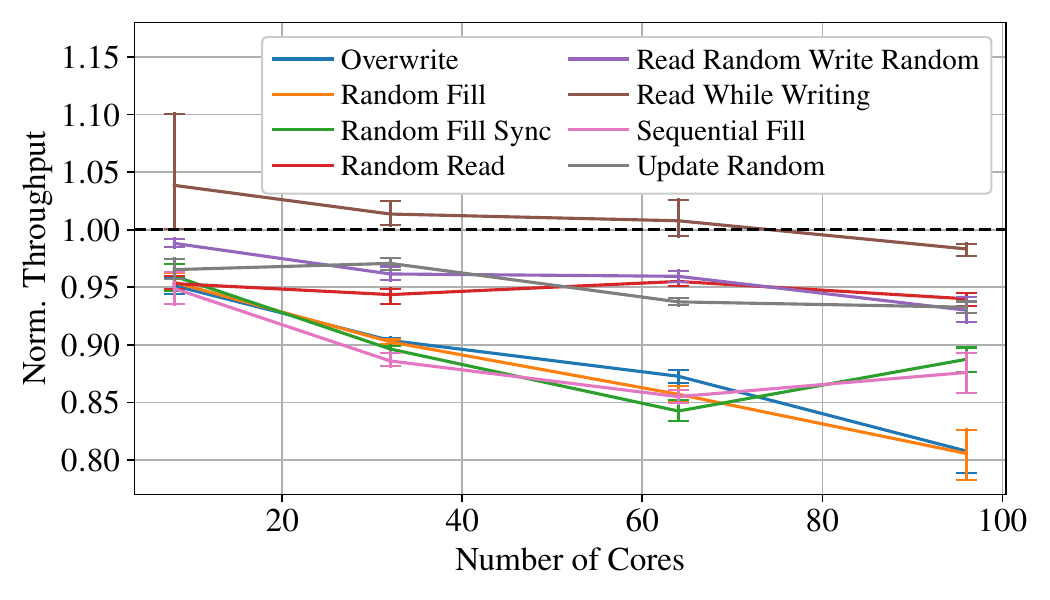}
        \caption{RocksDB: tests on different I/O patterns in database operations.}
        \label{fig:ampere-rocksdb}
    \end{subfigure}
    \begin{subfigure}{\linewidth}
        \centering
        \includegraphics[width=\linewidth]{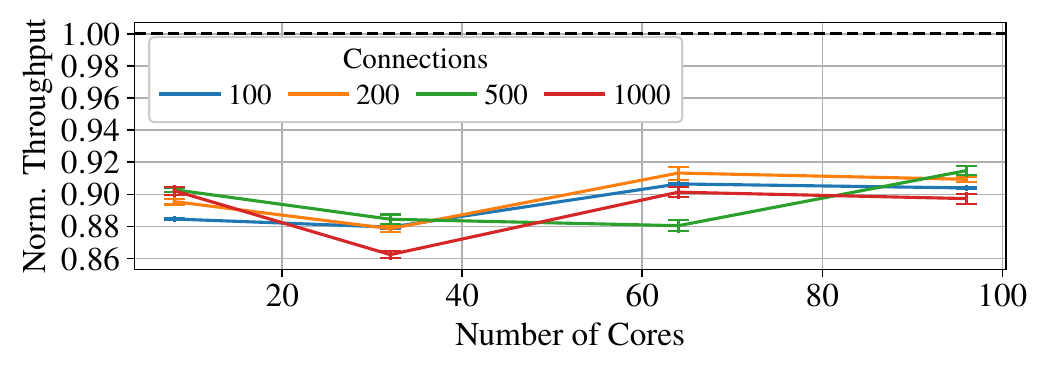}
        \caption{Nginx: tests on serving static files.}
        \label{fig:ampere-nginx}
    \end{subfigure}
    \begin{subfigure}{\linewidth}
        \centering
        \includegraphics[width=\linewidth]{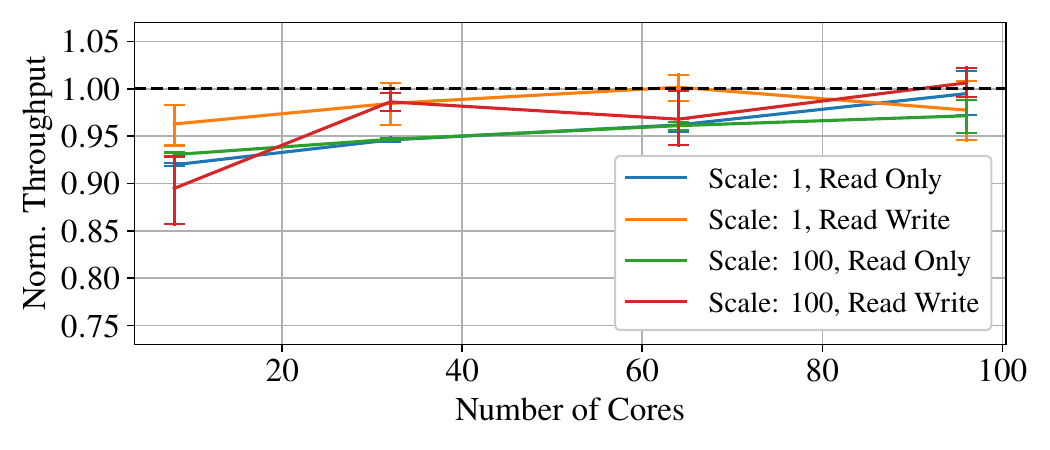}
        \caption{PostgreSQL: SQL queries on databases of different size.}
        \label{fig:ampere-postgres}
    \end{subfigure}

    \caption{MTE server workload performance on the AmpereOne \textbf{without the kernel patch} introduced in \S\ref{subsec:server}. MTE slowdowns for RocksDB, Nginx, PostgreSQL remain under \ampererocksdbmaxdrop. }
    \label{fig:ampere-server-bench-other}
\end{figure}

\begin{figure}[t]
  \centering

  \begin{subfigure}[b]{\linewidth}
    \centering
    \begin{minipage}{0.48\linewidth}
      \centering
      \includegraphics[width=\linewidth]{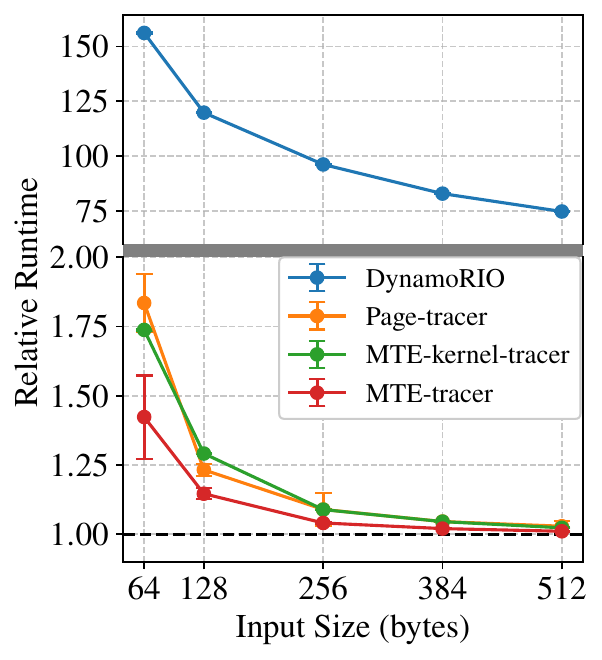}
      \subcaption*{RSA Sign}
    \end{minipage}\hfill
    \begin{minipage}{0.48\linewidth}
      \centering
      \includegraphics[width=\linewidth]{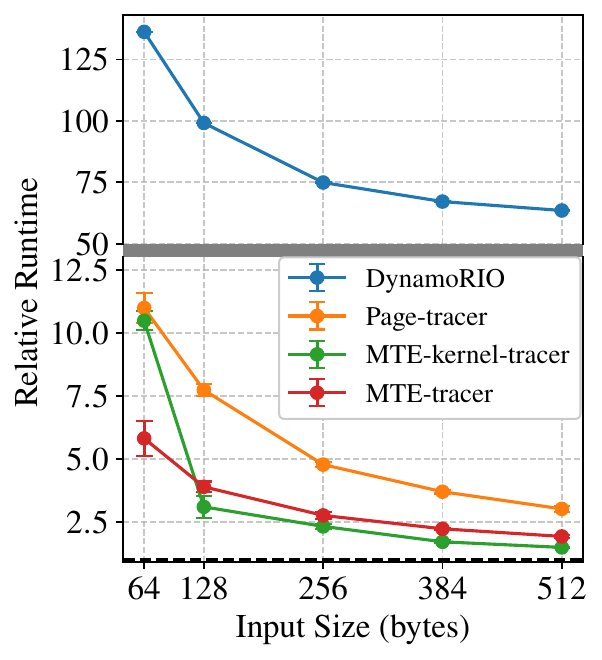}
      \subcaption*{RSA Verify}
    \end{minipage}
    \caption{AmpereOne}\label{fig:memtrace-ampere}
  \end{subfigure}

  \vspace{0.7em}

  \begin{subfigure}[b]{\linewidth}
    \centering
    \begin{minipage}{0.48\linewidth}
      \centering
      \includegraphics[width=\linewidth]{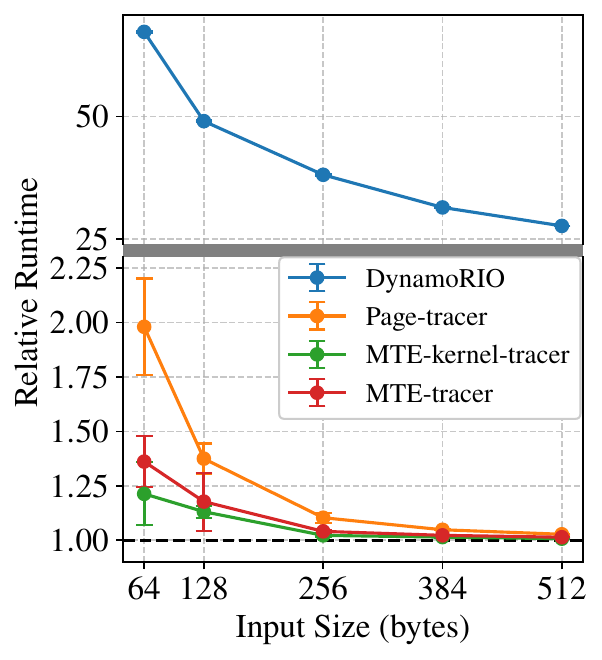}
      \subcaption*{RSA Sign}
    \end{minipage}\hfill
    \begin{minipage}{0.48\linewidth}
      \centering
      \includegraphics[width=\linewidth]{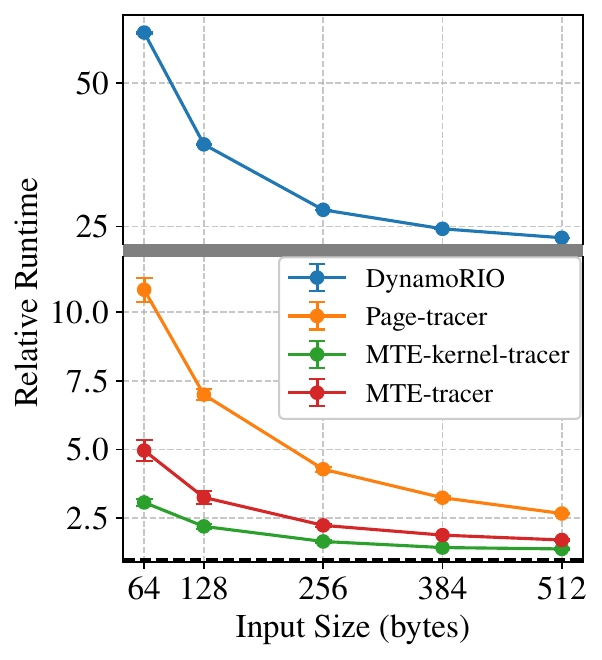}
      \subcaption*{RSA Verify}
    \end{minipage}
    \caption{Pixel 8 Little Core}\label{fig:memtrace-p8l}
  \end{subfigure}

  \vspace{0.7em}

  \begin{subfigure}[b]{\linewidth}
    \centering
    \begin{minipage}{0.48\linewidth}
      \centering
      \includegraphics[width=\linewidth]{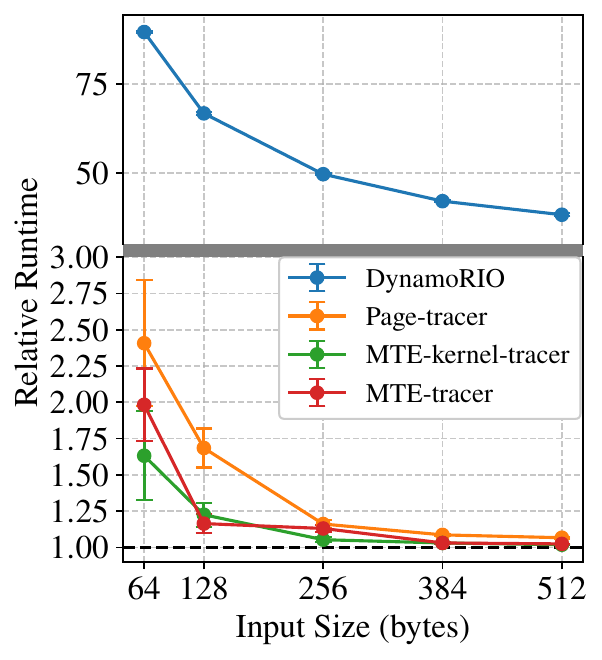}
      \subcaption*{RSA Sign}
    \end{minipage}\hfill
    \begin{minipage}{0.48\linewidth}
      \centering
      \includegraphics[width=\linewidth]{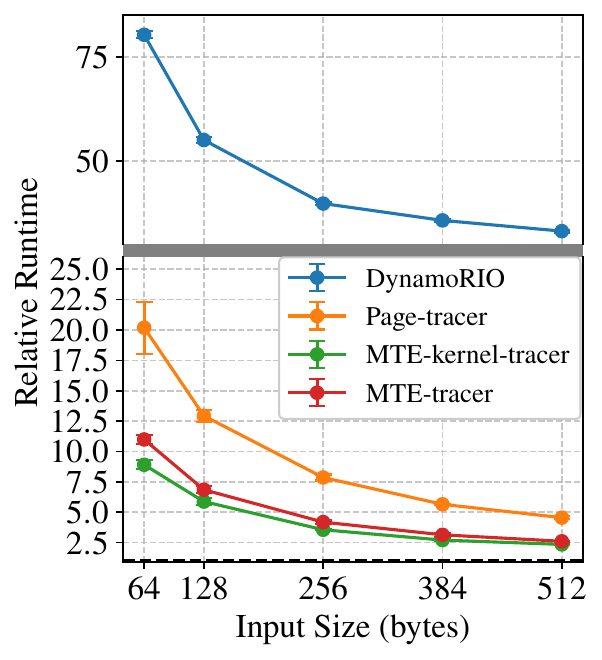}
      \subcaption*{RSA Verify}
    \end{minipage}
    \caption{Pixel 8 Big Core}\label{fig:memtrace-p8b}
  \end{subfigure}

  \caption{MTE-base and Page-based data tracing overheads on AmpereOne, Pixel 8 Little Core and Pixel 8 Big Core:
\textit{We compare the performance of tracing key material through OpenSSL using different tracers; the time taken is normalized to the baseline performance with no tracing enabled.
We see that MTE-based tracer is orders of magnitude faster than DynamoRio, and is 2 to 3 times faster than Page-tracer for small-sized inputs and the gap deceases as the buffer size increases.}
}
  \label{fig:memtrace-all}
\end{figure}

\begin{figure*}[t]
  \centering

  \begin{subfigure}[b]{0.95\textwidth}
    \centering
    \includegraphics[width=\linewidth]{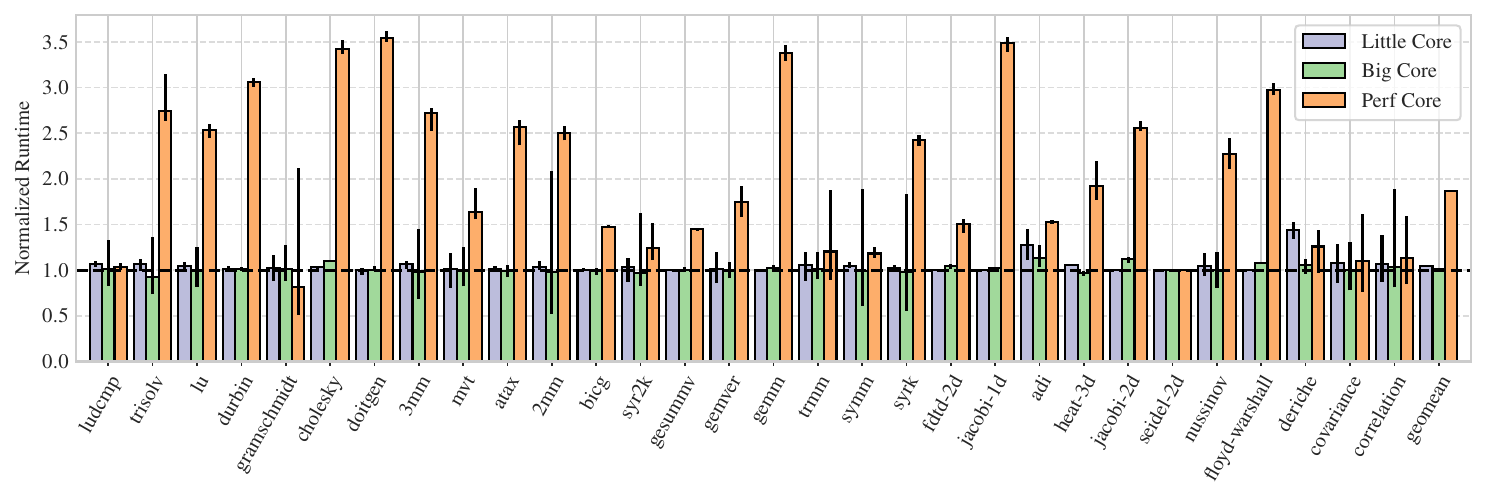}
    \caption{Pixel 8 cores.}
    \label{fig:polybench-pixel8}
  \end{subfigure}

  \vspace{0.7em}

  \begin{subfigure}[b]{0.95\textwidth}
    \centering
    \includegraphics[width=\linewidth]{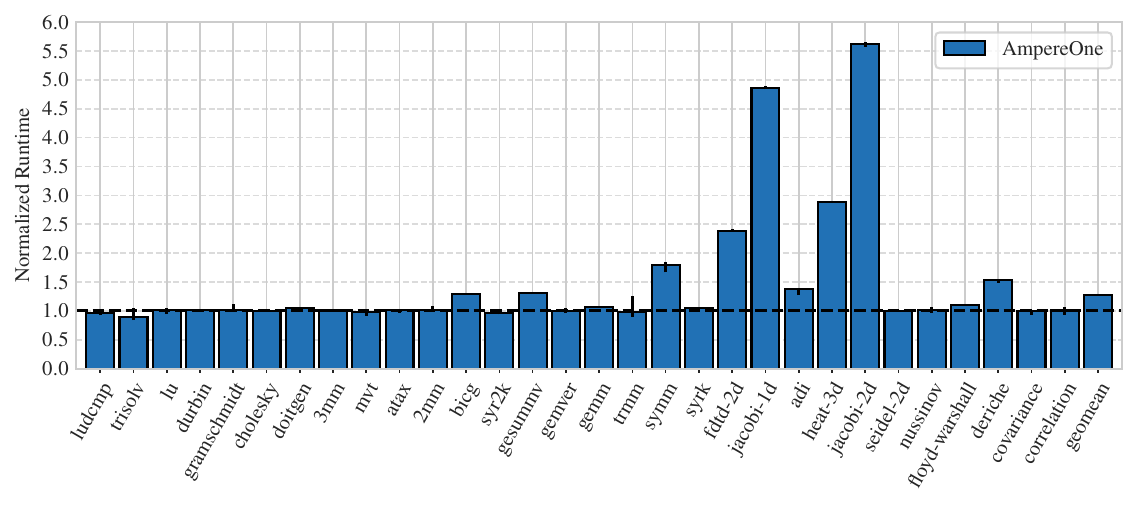}
    \caption{AmpereOne}
    \label{fig:polybench-ampere}
  \end{subfigure}

  \caption{
  The overheads of using ColorGuard with MTE SYNC on the PolyBench/C benchmark suite across platforms. 
  Colorguard eliminates the need for guard pages in SFI toolchains, but must tag all memory with a single tag for this. 
  This tagging generally results in low overheads for the Little and Big cores with geomeans of \polybenchlittlegeomean and \polybenchbiggeomean respectively, but has a high geomean overhead of \polybenchxgeomean in the performance core. 
  This means while ColorGuard can be useful in this domain, the amount of performance variation does not make it the ideal approach.}
  \label{fig:polybench-all}
\end{figure*}

}

\end{document}
\endinput